\documentclass[final]{agujournal2019}
\usepackage[hidelinks]{hyperref}
\usepackage{lineno}
\usepackage[inline]{trackchanges}
\usepackage{soul}
\usepackage{amsmath}
\usepackage{amssymb}
\usepackage{natbib}
\usepackage{mathtools}
\usepackage{caption}
\usepackage{subcaption}
\usepackage{float}
\usepackage{xcolor}
\usepackage{extpfeil}
\usepackage{booktabs}
\usepackage{tikz}
\usepackage[utf8]{inputenc}
\usepackage{rotating}
\usepackage{bm}

\usepackage{array}
\newcolumntype{L}[1]{>{\raggedright\let\newline\\\arraybackslash\hspace{0pt}}m{#1}}
\newcolumntype{C}[1]{>{\centering\let\newline\\\arraybackslash\hspace{0pt}}m{#1}}
\newcolumntype{R}[1]{>{\raggedleft\let\newline\\\arraybackslash\hspace{0pt}}m{#1}}

\newcommand{\commentout}[1]{}

\definecolor{codegreen}{rgb}{0,0.6,0}
\definecolor{codegray}{rgb}{0.5,0.5,0.5}
\definecolor{codepurple}{rgb}{0.58,0,0.82}
\definecolor{backcolour}{rgb}{0.95,0.95,0.92}

\definecolor{authora}{RGB}{105, 41, 197}
\definecolor{authorb}{RGB}{17, 146, 233}
\definecolor{authorc}{RGB}{0, 93, 93}
\definecolor{authord}{RGB}{159, 24, 83}
\definecolor{authore}{RGB}{250, 78, 86}
\definecolor{authorf}{RGB}{82, 4, 8}
\definecolor{authorg}{RGB}{25, 127, 56}
\definecolor{authorh}{RGB}{0, 45, 159}
\definecolor{authori}{RGB}{238, 83, 152}
\definecolor{authorj}{RGB}{178, 133, 0}
\definecolor{authork}{RGB}{0, 157, 164}
\definecolor{authorl}{RGB}{1, 39, 74}
\definecolor{authorm}{RGB}{138, 56, 0}
\definecolor{authorn}{RGB}{166, 110, 255}

\newcommand*{\authorcomment}[3][Crimson]{\textcolor{#1}{\textbf{\emph{#3}} }}

\newcommand*{\eliot}[1]{\authorcomment[authord]{}{#1}}
\renewcommand*{\eliot}[1]{}

\newcommand{\Upp}{U^{\prime \prime, \tau}}
\newcommand{\Uppt}{U^{\prime \prime, \tau + \delta \tau}}

\newcommand{\Vppt}{V^{\prime \prime, \tau + \delta \tau}}

\newcommand{\Tpp}{\Theta^{\prime \prime, \tau}}
\newcommand{\Tppt}{\Theta^{\prime \prime, \tau + \delta \tau}}
\newcommand{\Rpp}{\rho^{\prime \prime, \tau}}
\newcommand{\Rppt}{\rho^{\prime \prime, \tau + \delta \tau}}

\newcommand{\Up}{U^{\prime \prime}}
\newcommand{\Vp}{V^{\prime \prime}}
\newcommand{\Wp}{W^{\prime \prime}}
\newcommand{\Rp}{\rho^{\prime \prime}}
\newcommand{\Tp}{\Theta^{\prime \prime}}

\newcommand{\sfrac}[2]{\mathchoice
  {\kern0em\raise.5ex\hbox{\the\scriptfont0 #1}\kern-.15em/
   \kern-.15em\lower.25ex\hbox{\the\scriptfont0 #2}}
  {\kern0em\raise.5ex\hbox{\the\scriptfont0 #1}\kern-.15em/
   \kern-.15em\lower.25ex\hbox{\the\scriptfont0 #2}}
  {\kern0em\raise.5ex\hbox{\the\scriptscriptfont0 #1}\kern-.2em/
   \kern-.15em\lower.25ex\hbox{\the\scriptscriptfont0 #2}}
  {#1\!/#2}}
\newcommand{\half}{\sfrac{1}{2}}

\draftfalse

\journalname{JAMES}

\begin{document}

\title{ERF: Energy Research and Forecasting Model}

\authors{Aaron Lattanzi\affil{1}, Ann Almgren\affil{1}, Eliot Quon\affil{2}, Mahesh Natarajan\affil{1},   Branko Kosovic\affil{3}, Jeffrey Mirocha\affil{4},  Bruce Perry\affil{2}, David Wiersema\affil{4}, Donald Willcox\affil{1}, Xingqiu Yuan\affil{5}, Weiqun Zhang\affil{1}}

\affiliation{1}{Lawrence Berkeley National Laboratory}
\affiliation{2}{National Renewable Energy Laboratory}
\affiliation{3}{Johns Hopkins University}
\affiliation{4}{Lawrence Livermore National Laboratory}
\affiliation{5}{Argonne National Laboratory}

\correspondingauthor{Aaron Lattanzi}{amlattanzi@lbl.gov}

\begin{keypoints}
\item The Energy Research and Forecasting (ERF) code can leverage modern computational resources to simulate regional atmospheric flows.
\item ERF supports fully compressible and anelastic equation sets with moisture and terrain for atmospheric flow characterization.
\item ERF has been verified against benchmark data for a variety of idealized test cases.
\end{keypoints}

%
%

\begin{abstract}
High performance computing (HPC) architectures have undergone rapid development in recent years. As a result, established software suites face an ever increasing challenge to remain performant on and portable across modern systems. Many of the widely adopted atmospheric modeling codes cannot fully (or in some cases, at all) leverage the acceleration provided by General-Purpose Graphics Processing Units (GPGPUs), leaving users of those codes constrained to increasingly limited HPC resources. Energy Research and Forecasting (ERF) is a regional atmospheric modeling code that leverages the latest HPC architectures, whether composed of only Central Processing Units (CPUs) or incorporating GPUs. ERF contains many of the standard discretizations and basic features needed to model general atmospheric dynamics.
The modular design of ERF provides a flexible platform for exploring different physics parameterizations and numerical strategies. ERF is built on a state-of-the-art, well-supported, software framework (AMReX) that provides a performance portable interface and ensures ERF's long-term sustainability on next generation computing systems. This paper details the numerical methodology of ERF, presents results for a series of verification/validation cases, and documents ERF's performance on current HPC systems. The roughly 5x speed up of  ERF (using GPUs) over WRF (CPUs only) for a 3D squall line test case highlights the significance of leveraging GPU acceleration.
\end{abstract}

\section*{Plain Language Summary}
The Energy Research and Forecasting (ERF) model is a new code for simulating atmospheric flows that is able to efficiently model regional weather as well as local flow phenomena. ERF is built on a state-of-the-art, well-supported, software framework that allows ERF to run on computers from laptops through supercomputers.  In addition to its key uses for numerical weather prediction and studying atmospheric dynamics,
ERF is designed to provide a flexible computational framework for the exploration and investigation of different physics representations and solution strategies.


\section{Introduction} \label{sec:intro}

Computational modeling plays a major role in improving our fundamental understanding of atmospheric flow physics, with practical application to weather forecasting and insight into extreme weather events. 
The wide range of spatiotemporal scales present in such systems, from global to mesoscale to microscale, makes accurate modeling of atmospheric flows challenging.

Engineering applications tend to focus on the dynamics of the turbulent atmospheric boundary layer (ABL).
Research in these areas benefit from advances in high-performance computing (HPC) architectures, which can dramatically increase simulation throughput \citep{Donahue2024_SCREAM}, thereby enabling larger problem sizes and/or shorter times to solution. For example, studies of tropical cyclones can require large mesoscale domains (3,000~km $\times$ 3,000~km) with high-resolution nested microscale patches (80~km $\times$ 80~km) \citep{Stern2021_EstimatingRiskExtreme}. Flow within the microscale domain is typically calculated with large-eddy simulation (LES) and grid spacing $\Delta \sim \mathcal{O}(10)$~m \citep{Stern2021_EstimatingRiskExtreme, SanchezGomez2023_WindFieldsCategory}. However, an order of magnitude higher resolution ($\Delta \sim \mathcal{O}(1)$~m) is needed to capture surface winds during a hurricane \citep{Ma2024_LargeEddySimulation}, which pushes the limit of computational resources if meso- and microscales are considered simultaneously. These resolution requirements for LES also apply to the simulation of canonical ABLs, with stable boundary layers requiring $\Delta \sim \mathcal{O}(1)$~m \citep{Beare2006_GABLS1,Sauer2020_FastEddy,Wurps2020_GridResolutionRequirementsLargeEddy}, unless advanced turbulence models are used \citep{KosovicCurry2000,Basu2006_LargeEddySimulationStably,Zhou2011_LargeEddySimulationStable}. Increased computational throughput also enables the simulation of more comprehensive ensembles that characterize uncertainty
\citep{Ren2015_EnsembleMethodsWindSolar, Yan2022_UncoveringWindPower}. For example, over 50 simulations of the Weather Research and Forecasting (WRF) \citep{WRF:Skamarock} model were used to create a European wind resource atlas \citep{Hahmann2020_NEWA}; similar efforts have been undertaken with WRF in the U.S.
\citep{Draxl2015_WINDToolkit, Bodini2024_NOW23}. While existing numerical weather prediction (NWP) models, such as WRF, have seen widespread use in regional weather prediction and dynamic downscaling of mesoscale conditions to turbulence-resolving microscale flows \citep{Moeng2007_ExaminingTwoWayGrid, Mirocha2014_ResolvedTurbulenceCharacteristics}, there is currently no established community code that can provide all of these modeling capabilities and leverage graphics processing units (GPUs) on modern HPC systems \citep{Haupt2020_industry_workshop}. To address these needs, we have developed the open-source Energy Research and Forecasting (ERF) model \citep{ERF_joss}.

ERF incorporates many of the discretizations and algorithmic choices from WRF to provide a computationally efficient atmospheric simulation capability for mesoscale, idealized microscale, and mesoscale-to-microscale downscaled flows. However, it goes beyond the current WRF capabilities in three key ways: in performance portability across architectures; in the generality of the meshing strategies; and in providing an anelastic option as an alternative to the fully compressible equations. The most fundamental difference between the WRF and ERF dynamical cores is the terrain coordinate system; ERF utilizes height-following coordinates while WRF employs hybrid pressure-based coordinates. Furthermore, we emphasize that ERF is not a ``port'' of WRF to GPUs but rather an entirely new open-source C++ code base that incorporates existing algorithms and discretizations. We also acknowledge that there are numerous physics and microphysics options accessible through WRF that are not yet available through ERF. For example, radiation and land surface models have been integrated into ERF but are not yet verified; see discussion in Section~\ref{sec:conclusion} for more details. Additionally, data assimilation is not yet available in ERF. 

A summary of all global and regional atmospheric simulation codes is beyond the scope of this paper. Therefore, we briefly discuss {\it regional} simulation codes able to run on GPUs from at least one vendor. AceCAST \citep{AceCAST}, FastEddy \citep{Sauer2020_FastEddy, munoz-esparza_fasteddy_2022}, NUMA \citep{NUMA}, COSMO \citep{fuhrer_near-global_2018} and ClimateMachine \citep{Sridhar_2022} solve the compressible form of the equations. AceCAST is self-described as a ``modified implementation of the standard CPU-based WRF model'', and includes refactored WRF physics and dynamics modules that have been enabled to run on GPUs by way of CUDA and OpenACC directives. AceCAST is not open source and, according to recent documentation, only targets NVIDIA architectures. FastEddy is a large-eddy simulation (LES) code with warm moisture physics and a generalized coordinate system for terrain-following coordinates. However, unlike WRF and ERF, it uses a co-located spatial discretization, rather than an Arakawa C-grid.  Like AceCAST, FastEddy uses CUDA for GPU acceleration and thus targets NVIDIA GPUs only; COSMO uses OpenACC directives and also targets NVIDIA GPUs only. NUMA and ClimateMachine use a nodal discontinuous Galerkin (rather than finite volume) discretization and include adaptive mesh refinement through the use of p6est \citep{psixest}, an extension to p4est \citep{pforest}.

GRASP \citep{GRASP-Open} (built from DALES \citep{heus_formulation_2010, DALES2}), AMR-Wind \citep{Sharma2024_ExaWindOpensourceCFD} and MicroHH~\citep{microhh}  can solve the incompressible/Boussinesq or anelastic equation sets and target wind energy applications. MicroHH uses CUDA for GPU acceleration and thus targets NVIDIA GPUs only. AMR-Wind, also based on AMReX, is portable across multiple GPU architectures, but does not include terrain coordinates, grid stretching, moisture physics, or several other processes necessary for general atmospheric simulations.

While this is not a complete list, we are unaware of any other atmospheric modeling codes that run with NVIDIA (CUDA), AMD (HIP) and Intel (SYCL) GPUs, utilize adaptive mesh refinement in addition to grid stretching and terrain-following coordinates, and have the run-time option to switch between fully compressible and anelastic equation sets. We believe ERF is unique in this combination of capabilities.

In addition to the features above, ERF offers the capability to evolve Lagrangian particles in time. It is known that physical particles play a significant role in many atmospheric flows, such as pollutant and aerosol transport \citep{Alam_2008}, atmospheric chemistry \citep{Lin_2013}, Lagrangian cloud microphysics \citep{Morrison_2020} and wind-borne debris (from tornadoes or hurricanes) \citep{zhao_review_2021}.
ERF ``particles'' can represent physical particles or particle aggregates; they can also be used as more generic tracer quantities advected with the flow. Using the AMReX support for particles and particle-mesh operations, ERF particles can also interact with the flow field and other quantities stored on the mesh. We show an example of passive particle transport by a moist flow in Section~\ref{sec:vandv}.
 
The paper is organized as follows.  In Section~\ref{sec:goveqs}, we describe the governing equations for compressible and anelastic flows as well as the turbulence and microphysics models, physical forcings, and options for initial and boundary conditions.  In Section~\ref{sec:numerics},  we describe the numerical discretization of these equations, both spatial and temporal, and describe ERF's approach to mesh refinement. 
To validate ERF's numerics and showcase ERF's capabilities, we present a set of general atmospheric test cases
in Section~\ref{sec:vandv}.   In Section~\ref{sec:software}, we give some background about AMReX and present performance results.  Finally, we discuss current and future work in Section~\ref{sec:conclusion}.

\section{Governing equations} \label{sec:goveqs}

Two modes of operation are supported in the ERF dycore: in the first, the fully compressible fluid equations are solved on the mesh; in the second, the velocity field obeys an anelastic constraint and the density field is approximated with the hydrostatic density.  Subgrid-scale models for large-eddy simulation (LES), planetary boundary layer (PBL) parameterizations, and additional source terms including microphysics, simplified radiative source terms, and additional forcing terms that can be used with both modes are discussed further below.

For both formulations we first define a fixed-in-time base state (also referred to as background state or reference state) for pressure $p_0(z)$, density $\rho_0(z)$, dry potential temperature $\theta_{d_0}(z)$ and water vapor $q_{v_0}(z)$. This state is in hydrostatic equilibrium (HSE):
\[ \frac{d p_0}{dz} = - \rho_0 (z) g \]
where $g$ is the magnitude of the gravity vector ${\bf g}.$  Additionally the base state satisfies the equation of state (EOS) given below. For further detail about how the base state is constructed see Section~\ref{subsec:BCs}.  

\subsection{Fully compressible equation set} \label{subsec:goveqs_comp}

The compressible equations for the ERF model can be written as
\begin{align}
  \frac{\partial \rho_d}{\partial t} &= - \nabla \cdot (\rho_d \mathbf{u}), \label{eq:con} \\
    \frac{\partial (\rho_d \mathbf{u})}{\partial t} &= - \nabla \cdot (\rho_d \mathbf{u} \mathbf{u}) - \frac{1}{1 + q_v + q_c} ( \nabla p^\prime  - \delta_{i,3}\mathbf{B} ) - \nabla \cdot \boldsymbol{\tau} + \mathbf{F}_{u}, \label{eq:mom}\\
  \frac{\partial (\rho_d \theta_d)}{\partial t} &= - \nabla \cdot (\rho_d \mathbf{u}        \theta_d) + \nabla \cdot ( \rho_d \alpha_{\theta}\ \nabla \theta_d) + F_{\theta} + H_{n} + H_{p}, \label{eq:theta}\\ 
   \frac{\partial (\rho_d \boldsymbol{\phi})}{\partial t} &= - \nabla \cdot (\rho_d \mathbf{u} \boldsymbol{\phi}) + \nabla \cdot ( \rho_d \alpha_{\phi}\ \nabla \boldsymbol{\phi}) + \mathbf{F}_{\phi}, \label{eq:scalar}\\
   \frac{\partial (\rho_d \mathbf{q_{n}})}{\partial t} &= - \nabla \cdot (\rho_d \mathbf{u} \mathbf{q_{n}}) + \nabla \cdot (\rho_d \alpha_{q} \nabla \mathbf{q_{n}}) + \mathbf{F_{n}} + \mathbf{G_{p}}, \label{eq:nonprecip}\\
   \frac{\partial (\rho_d \mathbf{q_{p}})}{\partial t} &= - \nabla \cdot (\rho_d \mathbf{u} \mathbf{q_{p}}) + \partial_{z} \left( \rho_d \mathbf{w_{t}} \mathbf{q_{p}} \right) + \mathbf{F_{p}}. \label{eq:precip}
\end{align}
where $\mathbf{u} = \left[ u \;\; v \;\; w \right]$ is the fluid velocity vector and $p^{\prime} = p - p_{0}$ is the perturbational pressure. This equation set is supplemented by an equation of state (EOS),
 \begin{align}
 p = P_{00} \left( \frac{R_d \; \rho_{d} \; \theta_d \; (1 + \frac{R_v}{R_d} q_v)}{P_{00}} \right)^{\gamma},
 \end{align}
which allows us to diagnose pressure $p$ from the dry density $\rho_d$, dry potential temperature $\theta_d,$ and water vapor mixing ratio $q_v$.  Here $\gamma = c_p / (c_p - R_d),$  where $R_d$ and $c_p$ are the gas constant and specific heat capacity for dry air, respectively,  and $P_{00}$ is a reference value for pressure, set in ERF to be $1\times 10^{5}$ [Pa].
 
Here $\boldsymbol{\tau}$ is the stress tensor and $\mathbf{B}=-\left( \rho - \rho_{0}  \right)\mathbf{g}$ is the buoyancy force, which can also be approximated in ERF as $-\rho_0 (\theta - \theta_0) / \theta_0 \mathbf{g}$ or $-\rho_0 (T - T_0) / T_0 \mathbf{g}$ where $T=\theta_{d} \left(p/P_{00}\right)^{R_d/c_p}$ is the temperature and $T_0(z)=\theta_{d_0} \left(p_0/P_{00}\right)^{R_d/c_p}$ is the base state temperature. 
$\mathbf{F}_{u}$, $F_{\theta}$, and $\mathbf{F}_{\phi}$ are momentum, thermal, and scalar forcing terms (see Section~\ref{subsec:forcings}) and $\alpha_{i}$ is the diffusivity for scalar $i$.

The vector of transported scalars, $\boldsymbol{\mathbf{\phi}} = \left[ k \;\; \phi_{0} \;\; ... \;\; \phi_{n} \right]$, includes turbulent kinetic energy (TKE), $k$, along with arbitrarily many additional scalars, $\phi_{i}, i = 0, ..., n$. Source terms for each scalar are contained in $\mathbf{F}_{\phi}$ and further details on the turbulence models will be provided in Section~\ref{subsec:turbmodels}. ERF's ability to run with up to thousands of advected quantities allows it to serve as a core platform for atmospheric chemistry calculations where many chemical species must be transported. In general, the non-precipitating water mixing ratio vector $\mathbf{q_{n}} = \left[ q_v \;\; q_c \;\; q_i \right]$ includes water vapor, $q_v$, cloud water, $q_c$, and cloud ice, $q_i$, although some models may not include cloud ice; similarly, the precipitating water mixing ratio vector $\mathbf{q_{p}} = \left[ q_r \;\; q_s \;\; q_g \right]$ involves rain, $q_r$, snow, $q_s$, and graupel, $q_g$, though some models may not include these terms. The source terms for moisture variables, $\mathbf{F_{p}}$, $\mathbf{F_{n}}$, $\mathbf{G_{p}}$, and their corresponding impact on potential temperature, $H_{n}$ and $H_{p}$, and the terminal velocity, $\mathbf{w_{t}}$, are all specific to the employed microphysics model.

\subsection{Anelastic approximation} \label{subsec:Anelastic} 

The anelastic equation set we define in ERF replaces the equation of state with a divergence constraint,
\begin{equation}
    \nabla \cdot (\rho_0(z) \; \mathbf{u}) = 0. \label{eq:anelastic}
\end{equation}
%
In much of the literature, the anelastic constraint is viewed as resulting from approximating $\rho_d$ by the fixed-in-time $\rho_0$ in the continuity equation (Equation~\ref{eq:con}).  However, more generally the constraint derives from taking the Lagrangian derivative of the equation of state and approximating the pressure by $p_0$ (see, e.g. \citet{Durran}). This more general derivation does not require smallness of $\rho_d - \rho_0.$  For the present purposes we supplement this derivation with the assumption that $\rho_d$ is close to $\rho_0$ thus we can replace $\rho_d$ by $\rho_0$ in the governing equations, which makes the continuity equation redundant.  However, the separability of these assumptions (smallness of $p-p_0$ vs smallness of $\rho-\rho_0$) will allow us in the future to extend this option to more general applications where the density variations are non-negligible.

 When using the anelastic approximation we define the buoyancy force, $\mathbf{B}$,  for dry air as  $\rho_0 (\theta - \theta_0) / \theta_0 \mathbf{g};$  other source terms are as defined for the compressible case. In this equation set the pressure perturbation, $p'$, that enters the momentum equation is computed solely from the Poisson equation that results from the divergence constraint and does not enter the equation of state or contribute to any thermodynamic processes.  We note that if the domain height is sufficiently small, the base state has negligible variation in the vertical direction, thus $\rho_0$ is effectively constant in space and the anelastic equation set reduces to the incompressible flow equations.

\subsection{Turbulence models} \label{subsec:turbmodels}

ERF utilizes an expansion of the dynamic viscosity into molecular and turbulent components, $\mu = \mu_{\rm m} + \mu_{\rm t}$, when evaluating the fluid stress tensor
\begin{equation}
    \boldsymbol{\tau} = -2\left(\mu_{\rm m} + \mu_{\rm t}\right) \boldsymbol{\sigma}, \label{eq:stresstensor} 
\end{equation}
where $\boldsymbol{\sigma} = \mathbf{S} - \mathbf{D}$, $\mathbf{S} = 1/2 \; \left(\nabla \mathbf{u} + \nabla \mathbf{u}^{\intercal} \right)$ is the strain rate, $\mathbf{D} = 1/3 \; \left(\nabla \cdot \mathbf{u}\right) \mathbb{I}$ is the expansion rate, and $\mathbb{I}$ is the identity matrix. The molecular viscosity may be specified as a run-time constant while the turbulent viscosity is parameterized through various sub-grid scale (SGS) closures. The Smagorinsky \citep{LES:Smagorinsky,LES:Lilly} model and 1.5--order TKE model of \citet{LES:Deardorff} are available for LES descriptions while the MYNN 2.5 model of \citet{PBL:Mellor} and \citet{PBL:Nakanishi} is available for PBL descriptions. We note that PBL schemes, unlike eddy viscosity turbulence models, assume horizontal homogeneity on account of the coarse horizontal grid resolution used in the mesoscale regime and, therefore, only model vertical diffusion processes. Furthermore, while not considered here, PBL schemes may also be non-local, with fluxes depending on the atmospheric state throughout a column of air, and/or coupled to other models to account for microphysics and radiation. Finally, the turbulence models noted above are contained within the ERF C++ code base and run on GPUs.

\subsection{Cloud Microphysics} \label{subsec:micro}

Microphysics models describe the evolution of clouds, precipitation, and phase change in the atmosphere through statistical and physical modeling approaches. Currently, ERF includes only Eulerian models for the microphysics, but the particle capability in ERF allows for the development of efficient Lagrangian microphysics models in the future. Similar to the ERF dycore and turbulence models, all microphysics options listed below have been (re)written in C++ and run on GPUs.

The microphysics models in ERF may be broadly grouped into models that include only cloud vapor, cloud water and rain, and those that, in addition, include cloud ice, graupel and snow. The default model of the first type follows that of \citet{kessler_1969,klemp_simulation_1978}. The models of the second type include the Morrison model \cite{Morrison:2009} and the cloud resolving model of \cite{marat_2003}, which is akin to the classic Purdue Lin model \citep{lin_bulk_1983,rutledge_mesoscale_1984}. For each microphysics model noted above, the saturation pressure in ERF is computed from the $8^{\rm th}$ order polynomial correlation of \citet{flatau_polynomial_1992}. We note that the calculation of saturation pressure in WRF may vary with microphysics model but the Clausius-Clapeyron and Tetens equation are often utilized.

\section{Numerical Methodology} \label{sec:numerics}

Here we describe the spatial and temporal discretizations used to approximate the continuous governing equations in Section~\ref{sec:goveqs}. Additionally, discussions relevant to time step selection, initial/boundary conditions, and physical forcing terms are provided in the following sections.

\subsection{Temporal Discretization: Compressible} \label{subsec:TempDiscretization:comp}

When solving the fully compressible equations, ERF uses a low-storage third-order Runge-Kutta (RK3) scheme \citep{wicker_time-splitting_2002}, i.e. to solve
\begin{align}
    \frac{{\rm d}\mathbf{S}}{{\rm d}t} &= f\left(\mathbf{S}\right),
\end{align}
where $\mathbf{S}$ is the state vector, we take the following three steps:
\begin{align}
    \mathbf{S}^{*} &= \mathbf{S}^{n} + \frac{1}{3} f\left(\mathbf{S}^{n}\right) \Delta t, \\
    \mathbf{S}^{**} &= \mathbf{S}^{n} + \frac{1}{2} f\left(\mathbf{S}^{*}\right) \Delta t, \\
    \mathbf{S}^{n+1} &= \mathbf{S}^{n} + \;\; f\left(\mathbf{S}^{**}\right) \Delta t.
\end{align}
Here the superscript $n$ refers to the previous timestep. The third order accuracy of this RK3 scheme for linear advection in ERF has been verified through a convergence study.

As in WRF, the default timestepping scheme includes semi-implicit acoustic substepping within each Runge-Kutta stage although ERF provides options to alternatively use no substepping or explicit substepping.  For the semi-implicit substepping we closely follow the scheme as written in Section 2 of \citet{klemp_conservative_2007}; those equations are included in \ref{app:C}.  Here we give a high-level description but point the interested reader to \citet{klemp_conservative_2007} for further explanation and motivation for the design choices used here. We note that Equations~\ref{eq:scalar}--\ref{eq:precip} do not directly participate in the substepping; these variables are advanced as if there is no substepping, with the exception that the normal momenta used to define the advective fluxes, $(\rho_d \mathbf{u} \cdot \mathbf{n)}$ (where $\mathbf{n}$ is the normal to that face), is the time average over the RK stage of the intermediate $(\rho_d \mathbf{u} \cdot \mathbf{n})$ values used to advance $\rho_d$ each substep. Thus, algorithmically, in each stage, the solution of Equations~\ref{eq:scalar}--\ref{eq:precip} must always take place after the substepping is complete.

Without substepping, we compute the source term $f(\mathbf{S})$ in each RK stage, multiply it by the associated timestep ($\Delta t / 3$, $\Delta t / 2$, $\Delta t$ for the first, second and third stage, respectively), then add that contribution to the previous time solution $\mathbf{S}^{n}.$

When substepping, we define a new set of variables that represent perturbations from the solution at the RK stage where we have most recently computed $f(\mathbf{S})$. The (five) governing equations for the evolution of density, momenta and potential temperature perturbations are provided in \ref{app:C}. In these equations, we use a linearization of the equation of state (valid for small departures from the solution at the previous RK stage) to express the pressure gradient in terms of the gradient of potential temperature.  We also linearize the advective terms in each equation but leave the source terms unchanged. This results in five linear equations with five unknowns.

For both the explicit and semi-implicit substepping algorithms, we first update the horizontal momenta using pressure gradient terms evaluated at the previous substep and predicted slightly forward in time; this is the divergence damping approach described in \citet{klemp_conservative_2007}. When using explicit substepping, we also update the vertical momentum explicitly, then use the updated momenta to compute the updated density and potential temperature. By contrast, when doing semi-implicit substepping, we reduce the (three) equations for vertical momentum, density and potential temperature to a single equation for vertical momentum in which all horizontal differences are computed explicitly but all vertical differences are treated semi-implicitly, resulting in a tridiagonal solve for each column.  Due to linearization, coefficients in the tridiagonal matrix do not depend on the substepped solution and thus can be computed once per RK stage.

\subsection{Temporal Discretization: Anelastic} \label{subsec:TempDiscretization:anel}

When solving the anelastic equations, ERF uses a second-order strong stability-preserving Runge-Kutta (SSP-RK) scheme \citep{ShuOsher88} with no acoustic substepping:
\begin{align}
    \mathbf{S}^{*} &= \mathbf{S}^{n} + f\left(\mathbf{S}^{n}\right) \Delta t, \\
    \mathbf{S}^{n+1} &= \mathbf{S}^{n} + \frac{1}{2} \; \left(  
    f(\mathbf{S^n}) + f(\mathbf{S^*}) \right) \Delta t.
\end{align}
We note this is the same scheme employed by the PyCLES solver in \cite{PresselEtAl2015}.
To enforce the anelastic constraint in each Runge-Kutta stage, we first perform the update in time without considering the constraint to define a provisional velocity field---e.g., $\mathbf{u}^{*,*}$ in the first stage $\mathbf{u}^{n+1,*}$ in the second.  This update explicitly solves Equation~\ref{eq:mom} using a lagged pressure pertubation, $p^{\prime}$.  After the update, we project the velocity field by solving, e.g., in the case of the first Runge-Kutta stage,  
\begin{equation}
    \Delta t \; \nabla \cdot (\nabla \psi)  = \nabla \cdot (\rho_0 \mathbf{u}^{*,*}),
\end{equation}
and setting
\begin{eqnarray*}
    \mathbf{u}^* &=& \mathbf{u}^{*,*} - \frac{\Delta t}{\rho_0} (\nabla \psi),  \\
    p^\prime &=& p^\prime + \psi.
\end{eqnarray*}
An analogous procedure is followed in the second RK step.
ERF provides several options for solving the Poisson equation: geometric multigrid, Fast Fourier Transforms (FFTs) and preconditioned GMRES.

\subsection{Choosing the Timestep} \label{subsec:dt} 

The solver timestep can be specified by the user or computed dynamically at each timestep based on the user-specified Courant--Friedrich--Lewy (CFL) number which we call ${\rm CFL}_{\rm user}$ here ---i.e., adaptive time stepping. For the compressible equations, the adaptive timestep calculation uses the acoustic CFL constraint, i.e. if not using the semi-implicit substepping we define
\[ \Delta t = {\rm CFL}_{\rm user} \; {\rm min} \; \left( \frac{\Delta x}{ |u| + c}, \; \frac{\Delta y}{|v| + c}, \; \frac{\Delta z}{|w| + c} \right); \]
when using semi-implicit acoustic substepping the vertical mesh spacing does not appear in the time step calculation, i.e.
\[ \Delta t = {\rm CFL}_{\rm user} \; {\rm min} \; \left( \frac{\Delta x}{|u| + c}, \; \frac{\Delta y}{|v| + c} \right). \]
Note that the minimum value is computed over all cells at this level of resolution in the domain, and for variable mesh spacing, the local value of $\Delta z$ is used for each cell.

For the anelastic equations, the adaptive timestep calculation uses the advective CFL constraint, which is determined by the fluid speed rather than the sound speed and thus allows much larger timesteps, i.e. 
\[ \Delta t = {\rm CFL}_{\rm user} \; {\rm min} \; \left( \frac{\Delta x}{|u|}, \; \frac{\Delta y}{|v|}, \; \frac{\Delta z}{ |w|} \right); \]

When using acoustic substepping for the compressible equations, the size of the acoustic substep can be specified by the user, dynamically computed using a user-specified "substepping CFL number", or calculated as a fixed fraction of the total timestep (with the constraint that the full timestep must be an even multiple of the substepping timestep). 
If the substep size is not specified by the user, ERF's default behavior is to utilize 6 substeps per timestep, which matches WRF. 
Both WRF and ERF default to using just one substep in the first RK stage since \citet{wicker_time-splitting_2002} showed that using only one substep in the first RK stage had no noticeable loss of accuracy or stability.

We note that because acoustic substepping does not involve all the variables, and solving it does not require re-computation of source terms or evaluation of the equation of state, one substep is considerably computationally cheaper than one RK stage.  Thus, substepping can be viewed as a way to reduce total computational cost by enabling a larger overall timestep $\Delta t$.  The selection of the vertical direction to be treated implicitly reflects that the mesh spacing in the vertical is often smaller than the mesh spacing in the horizontal, thus this implicitness removes the vertical mesh spacing from the CFL time step limitation.

\subsection{Spatial Discretization and Terrain Coordinates} \label{subsec:SpatialDiscretization} 

The spatial discretization in ERF uses the classic Arakawa C-grid with
scalar quantities at cell centers and normal velocities at cell faces.
Simulations over complex topography use a terrain-following, height-based, vertical coordinate \citet{klemp_conservative_2007,sullivan_large-eddy_2014}; see \ref{app:A} and \ref{app:B} for further details. When terrain-following coordinates are used, the surface topography is specified at nodes (cell corners) either analytically or through parsing a text file.  Grid deformation (i.e., the change in the grid due to the presence of non-flat terrain) occurs only in the vertical direction. Metric terms resulting from mesh deformation are included in the governing equations in the same manner as \citet{sullivan_large-eddy_2014}.

As in many atmospheric modeling codes, variable mesh spacing in the vertical direction is allowed with or without terrain. The heights of each level can be parsed from a text file as ``z levels'' (as in WRF), or calculated at run-time given an initial mesh spacing at the bottom surface and a specified growth rate.  In the presence of non-flat terrain, the mesh is smoothed from the specified terrain at the bottom of the computational domain to a flat surface at the top of the domain. Three smoothing approaches are offered in ERF: Basic Terrain Following (``BTF''), in which the influence of the terrain decreases linearly with height;  Smoothed Terrain Following (``STF''), in which small-scale terrain structures are progressively smoothed out of the coordinate system as height increases; or Sullivan Terrain Following (``Sullivan''), where the influence of the terrain decreases with the cube of height. The BTF and Sullivan smoothing may be written in a canonical form
\begin{align}
    z &= \zeta + \left(1 - \frac{\zeta}{z_{t}}\right)^{\beta} h(x,y),
\end{align}
where $\zeta$ is the vertical coordinate in computational space, $z_{t}$ is the physical height of the top of the domain, $h(x,y)$ is the height of the terrain at $\zeta=0$, and $\beta$ is a relaxation parameter. One obtains the BTF method with $\beta=1$ and the Sullivan method with $\beta=3$. For the STF method, we refer the interested reader to \citet{Klemp_2011}.

Additionally, ERF includes the capability to apply several common isotropic map projections (e.g., Lambert Conformal, Mercator); the inclusion of map scale factors in the compressible equations and acoustic substepping are discussed in \ref{app:A} and \ref{app:C}, respectively.

\subsection{Mesh Refinement} \label{subsec:MeshRefinement}

ERF also allows both dynamic and static mesh refinement with sub-cycling in time at finer levels of refinement.  Arbitrary integer refinement ratios are allowed although typically ratios of 2, 3 or 4 are used; refinement can also be anisotropic, allowing refinement in one coordinate direction but not another. We utilize two-way coupling, in which the coarse solution is used to provide Dirichlet boundary conditions for the fine solution and the fine solution is averaged down onto the coarser level.  In addition, at the coarse-fine interface, we reflux all advected scalars to ensure conservation; refluxing involves the replacement of fluxes on a coarse mesh by the space-and-time-averaged fluxes on the fine mesh, such that conservation in the multilevel case matches the single-level simulation. 

Data at each level in ERF are defined on disjoint logically rectangular regions of the domain known as patches (or grids or boxes).
The specification of where and by how much parts of the domain should be refined is very flexible.  The simplest approach is to specify one or more 3D logically rectangular regions analogous to WRF ``domains."   Alternatively or additionally, a user can specify a set of criteria for individual cells $(i,j,k)$, such that if the criteria are met by the data in that cell, that location is tagged for refinement.  ERF uses a ``clustering" routine \citep{bergerRigoutsos:1991} that defines a union of disjoint patches that contain the tagged cells and meet a user-specified grid efficiency criterion (that above a given fraction of the cells in the refined region must have been the originally tagged cells).
For parallel performance ERF requires at least one patch per MPI rank (thus a single rectangular region might be broken into multiple patches if run on multiple ranks), but 
unlike WRF, ERF does not require exactly one patch per MPI rank, thus allowing much more general
domain decomposition at all levels. 

To compute stencil operations in each patch we must provide data in ``ghost cells" (also known as ``halo cells") outside of the box itself.    These values are filled in one of three ways.  First, if the ghost cell lies outside the physical domain, physical boundary conditions are used to define values there. In the interior, if fine data from another patch is available, that is always used. If the ghost cells are only fillable from coarser data, we
conservatively interpolate in space and time. This interpolation scheme constructs local slopes in each cell / on each face to ensure that if the fine values were averaged down, their average would be identical to the coarse value in that cell / on that face. We also interpolate the normal momentum at the coarse-fine interface itself; this ensures mass conservation since the normal momentum is in fact the flux for the density field.  In order to ensure that the fine momentum on the coarse-fine boundary stays consistent with the interpolated coarse values throughout a fine timestep, we also interpolate the source term for the normal momentum on the coarse-fine interface.   When using the anelastic approximation, this ensures that the computation of the updates to the fine momentum do not use any pressure perturbation values from the coarser level, which are not synchronized between levels.

\subsection{Advection Schemes}

In ERF, the advective fluxes are obtained by multiplying the normal momentum, at the flux location, by the interpolated primitive variable (computed from one of the schemes below). When updating cell-centered quantities, the normal momentum is located at the flux location (cell face). When updating momenta, spatial averaging is utilized to obtain the normal momentum at the flux location. In the presence of complex terrain, we use $(\rho \Omega)$ rather than $(\rho w)$ on the vertical faces, where $\Omega$ is the velocity component normal to the face.  

Default interpolation methods in ERF include second- through sixth-order stencils, which include both centered difference and upwind schemes. For variable $q$ in direction $m$, the centered schemes (even) and  upwind schemes (odd) are
\begin{align}
    \left. q_{m - \frac{1}{2}} \right|^{2nd} &= \frac{1}{2}\left( q_{m} + q_{m - 1} \right),  \\
    \left. q_{m - \frac{1}{2}} \right|^{4th} &= \frac{7}{12}\left( q_{m} + q_{m - 1} \right)  
        - \frac{1}{12}\left( q_{m + 1} + q_{m - 2} \right), \\
    \left. q_{m - \frac{1}{2}} \right|^{6th} &= \frac{37}{60}\left( q_{m} + q_{m - 1} \right)
        - \frac{2}{15}\left( q_{m + 1} + q_{m - 2} \right) 
        + \frac{1}{60}\left( q_{m + 2} + q_{m - 3} \right), \\
    \left. q_{m - \frac{1}{2}} \right|^{3rd} &= \left. q_{m - \frac{1}{2}} \right|^{4th}
        +  \frac{\text{sign}(u_m)}{12} \left[ \left(q_{m+1} - q_{m-2} \right) 
        - 3\left(q_{m} - q_{m-1} \right)  \right], \\
    \left. q_{m - \frac{1}{2}} \right|^{5th} &= \left. q_{m - \frac{1}{2}} \right|^{6th}
        -  \frac{\text{sign}(u_m)}{60} \left[ \left(q_{m+2} - q_{m-3} \right)
        - 5\left(q_{m+1} - q_{m-2} \right)
        + 10\left(q_{m} - q_{m-1} \right)  \right].
\end{align}

Advection of cell centered scalars may also be computed with
third-, fifth-, or seventh-order weighted essentially non-oscillatory (WENO) schemes. Currently, ERF supports the classic WENO-JS \citep{jiang_efficient_1996}, WENO-Z \citep{borges_improved_2008}, and an improved third-order WENO-MZQ3 \citep{kumar_new_2023} method that preserves order near critical points. For the sake of brevity, we again refer the interested reader to the cited works for more details on the WENO schemes.

The horizontal and vertical interpolation operators may differ when the centered and upwind methods are utilized. However, when using a WENO method, the same scheme must be applied to all directions. Additionally, a monotonicity preserving third-order slope limited scheme is available \citep{hundsdorfer_positive_1995}. With this scheme, monotonicity is achieved with a flux limiter, $\phi(r)$, that is bounded such that $\phi(r) \geq 0$ and first-order upwind is obtained when $\phi(r) = 0$, where $r$ is the slope ratio. It is noted that the WENO schemes and monotonicity preserving switch are the primary efforts to directly control oscillations in the solution variables. For each of the schemes above, the observed order accuracy has been verified to match the theoretical order of accuracy for linear advection.

\subsection{Forcing and Sources} \label{subsec:forcings}

Physical forcings available in ERF comprise the standard source terms for atmospheric modeling: Coriolis and geostrophic forcing; Rayleigh damping and sponge layer(s); subsidence; simplified radiative thermal sources; and solution nudging towards a prescribed input sounding. General time- and height-varying tendencies may be included through user-defined functions.  


Removal of short-wavelength oscillations, $\mathcal{O} \left(2 \Delta x \right)$, can be crucial to maintaining numerical stability. To control numerical dissipation and high frequency fluctuations, ERF includes optional source terms from a sixth-order numerical diffusion operator \citep{xue_high-order_2000}. By default, the sixth-order diffusion is not active and, unless otherwise stated, was not utilized to generate results herein.

When solving the compressible equations with acoustic substepping, the user has the option to specify whether individual contributions to the source terms should be evaluated at the start of each Runge-Kutta stage or at the start of each acoustic substep, depending on the stiffness of the specific term.

\subsection{Initial and Boundary Conditions} \label{subsec:BCs}

For realistic problems, the initial data and base state may be read from {\tt met\_em} files generated by the WRF Preprocessing System (WPS) \citep{WRF:Skamarock} or from  the {\tt wrfinput} file generated during a WRF initialization, both of which are in NetCDF format. For idealized problems, the initial data and base state may be constructed from 1-D input sounding data or specified by the user. Custom perturbations, which are separately specified by the user, can be added to the initial and background state.  When constructing the background (HSE) state during initialization, we use a Newton--Raphson approach to solving the non-linear root finding problem that stems from requiring that the density, pressure and potential temperature satisfy both the hydrostatic balance and the equation of state. Users have the option to define a dry or moist background state. For ideal simulations based on an input sounding (as in WRF), the background state is initialized in HSE starting from the surface pressure, potential temperature, and water vapor mixing ratio; the Newton--Raphson approach is used to match the input profiles of potential temperature and water vapor mixing ratio at every height.

Lateral boundary conditions in ERF can be specified for idealized simulations as periodic, inflow, outflow, or open. More realistic conditions include the use of time-varying values read in from external files, such as the {\tt met\_em} files generated by the WPS or {\tt wrfbdy} files generated during a WRF initialization. If boundary files are used, ERF allows for nudging of the solution state, in a manner similar to WRF, towards the boundary data. ERF also has the option to run precursor simulations where  planes of data are saved at specified times and later read in as boundary data, analogous to the {\tt wrfbdy} files.

The bottom surface boundary condition can be specified as a simple (slip or no-slip) wall or by specifying surface momentum and scalar fluxes using Monin-Obukhov similarity theory (MOST) \citep{MOST:monin,MOST:van}. When utilizing MOST, the surface roughness, $z_{0}$, may be specified as a constant, read from a file, or dynamically computed from the Charnock \citep{charnock_wind_1955} or shallow water \citep{jimenez_need_2018} formulation. Time-varying land or sea surface temperatures may also be specified in conjunction with MOST.

\section{Verification \& Validation} \label{sec:vandv}

In Section~\ref{subsec:atmosphere} we first consider simulations relevant to general atmospheric flows. Sections~\ref{subsec:abl}--\ref{subsec:terrain} then consider microscale ABL and flow over terrain, respectively. In Section~\ref{subsub:dc} we explicitly compare anelastic and compressible results, with and without mesh refinement. However, in the interest of space, the remaining simulation results shown here were all run at a single level with the compressible formulation. Unless otherwise stated, the $3^{\rm rd}$ order advection scheme is utilized in the horizontal and vertical directions and compressible simulations employ acoustic substepping with 4-6 substeps per full timestep.

For a summary of the ERF simulation configurations considered in this section, we refer the reader to \ref{app:D} and Table~\ref{tab:BigObnoxiousThing}. Here, we briefly describe the organization of Table~\ref{tab:BigObnoxiousThing}. The domain extents describe the height of the domain in the 1-D single-column model (SCM) case; the length and height of the domain for the 2-D cases; and the streamwise, lateral, and vertical extents of the domain for the 3-D cases. Parenthetical quantities are derived from other inputs and have been included for completeness. Initial fields are either uniform or vary with height based on an input sounding. Perturbations were superimposed on the initial fields to model the specific mesoscale problem of interest or to help initiate turbulence in the microscale. For the canonical, non-neutral ABL simulations, MOST specifies the surface kinematic heat flux for the BOMEX case and the surface temperature for the GABLS cases. Both Smagorinsky and Deardorff 1.5 order TKE turbulence closures have been utilized; with Deardorff, the model coefficients follow \citet{Moeng1984}. Several cases (as noted in Table~\ref{tab:BigObnoxiousThing}) used Rayleigh damping to attenuate waves in the free atmosphere. When applicable, the damping strength, an inverse timescale, is provided and we note that ramping with a sine-squared function from 0 at the bottom of the damping layer to the specified value at the top was employed.

\subsection{Mesoscale Dynamics} \label{subsec:atmosphere}

Generally speaking, realistic atmospheric flows involve a variety of complex physical processes that are strongly coupled to the flow field. The idealized test cases considered here exercise different physics that are relevant to realistic problems---e.g., buoyant transport induced by thermal or moisture effects. In the following subsections, ERF is compared to benchmark results for problems involving moist and dry bubbles, as well as convective cells with precipitation.

\subsubsection{Density Current} \label{subsub:dc}

Density currents are a crucial feature in a variety of atmospheric phenomena such as thunderstorms, sea breezes, and cold front passages \citep{Droegemeier_1987}. The density current test case \citep{Straka_1993, Xue_2000_ARPS, Skamarock_2012} is a well-defined benchmark to assess numerical models that simulate atmospheric flows. The test case consists of a cold bubble that descends to the ground due to negative buoyancy, in a quiescent, hydrostatic background.

The initial condition consists of a cold bubble characterized by a temperature decrement given by
\begin{eqnarray}\label{eqn:SL_Delta_theta}
\Delta T = 
\begin{cases} 
-15\Bigg[\cfrac{\cos(\pi L) + 1}{2}\Bigg] & L<1 \\
0 & L > 1,
\end{cases}
\end{eqnarray}
where 
\begin{equation}
L = \sqrt{\Bigg(\frac{x - x_c}{x_r}\Bigg)^2 + \Bigg(\frac{z - z_c}{z_r}\Bigg)^2},
\end{equation}
with $x_c = 0.0$ m, $z_c = 3\times 10^3$ m, $x_r = 4\times 10^3$ m, and $z_r = 2\times 10^3$ m. Constant diffusivities of $\nu$ = $\alpha_{\theta}$ = 75 m$^2$/s are employed; no turbulence model is used.  Due to the symmetry of the initial conditions and the domain, here we run only half the domain with a symmetry boundary condition at $x=0$. The initial coarse grid spacing is $(100.0, \, 100.0)$~m with a timestep of $1.0$ s.
To assess grid convergence, we also evaluate medium and fine resolution grids with $50$~m and $25$~m grid spacing, respectively. The timestep size in these cases is reduced by the same factor as the grid spacing. Furthermore, we evaluate the effectiveness of AMR for a single level of refinement. In the AMR configuration, the grids were adaptively refined by a factor of four such that the background grid spacing matches the coarse setup and the refined region matches the fine setup.

\begin{figure}
     \centering
     \includegraphics[width=0.99\textwidth]{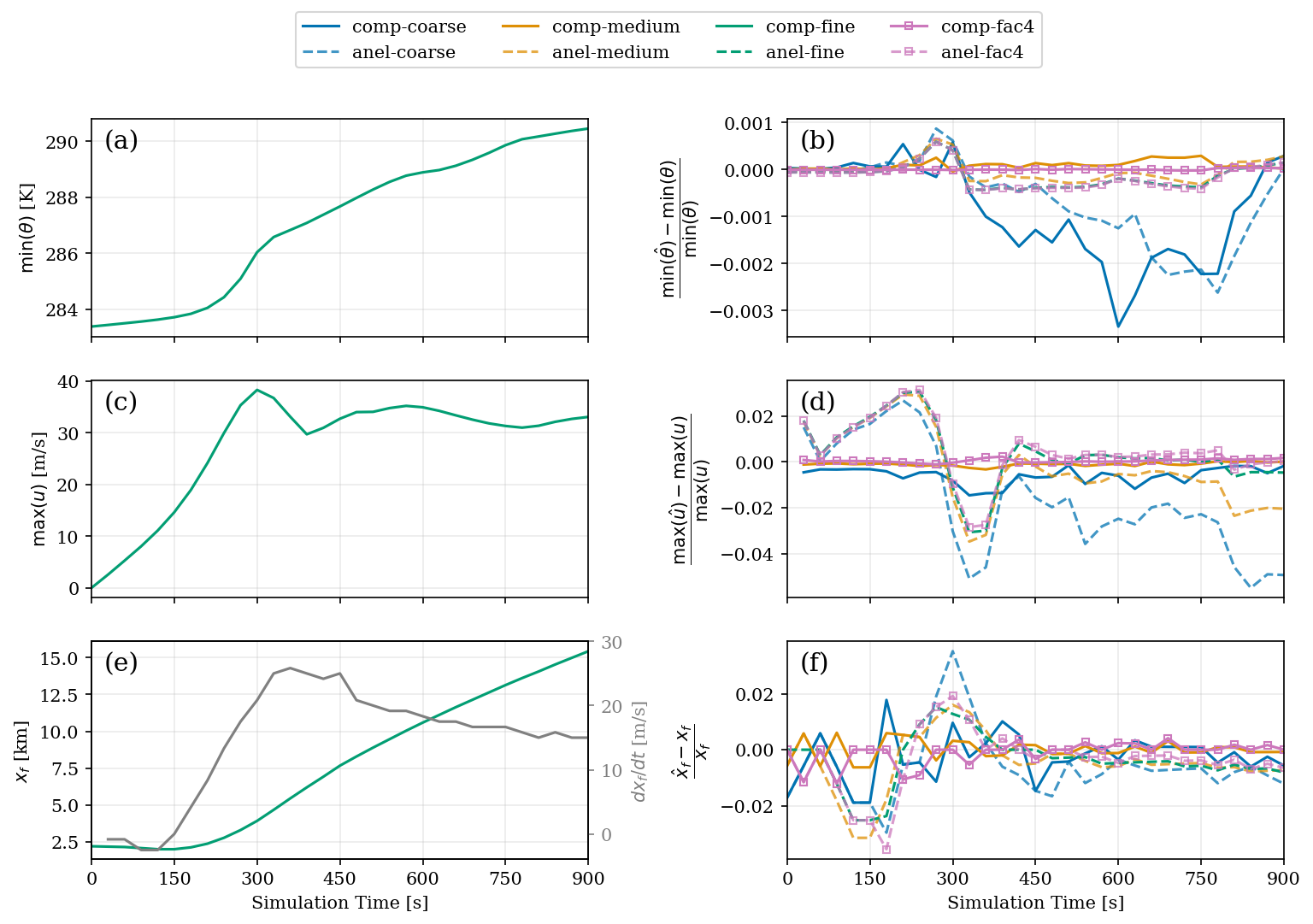}
     \caption{Density current time histories of minimum potential temperature (a,b), maximum outflow velocity (c,d), and front position (e,f) for compressible (``comp'') and anelastic (``anel'') simulation modes with coarse, medium, fine, and AMR grid configurations. The right panels show the normalized error of each simulated quantity (with hat symbols) relative to the compressible, fine-grid result in the left panels.}
     \label{fig:dc_time_hist}
\end{figure}

As the simulation proceeds, the cold bubble descends due to negative buoyancy, hits the bottom surface, and spreads laterally. Kelvin-Helmholtz shear instability results in the formation of vortices as the bubble propagates away from the starting position, towards the right in our symmetric domain. 
Figure~\ref{fig:dc_time_hist} shows the evolution of several bulk quantities of interest. Our reference solution (compressible mode, fine grid) shows temperature recovery as the initial cold bubble is deformed and diffused by the vortices. After 900~s, the minimum potential temperature is $290.5$~K along the surface near the front of the density current (Fig.~\ref{fig:dc_time_hist}(a)), which agrees with solutions in literature \citep{Skamarock_2012}. The outflow wind speed reaches a maximum at 300~s (38.3~m/s, Fig.~\ref{fig:dc_time_hist}(c)), which exceeds 30~m/s as described in \citet{Straka_1993}; between 390--600 s, the outflow wind speed ramps up to 35.2~m/s \citep[close to 36 m/s reported by][]{Straka_1993}. The front of the density current reaches a distance of 15.4~km \citep[Fig.~\ref{fig:dc_time_hist}(e), c.f.][]{Xue_2000_ARPS} with a speed of 15~m/s \citep[c.f.][]{Straka_1993}.

\begin{figure}
     \centering
     \includegraphics[width=0.99\textwidth]{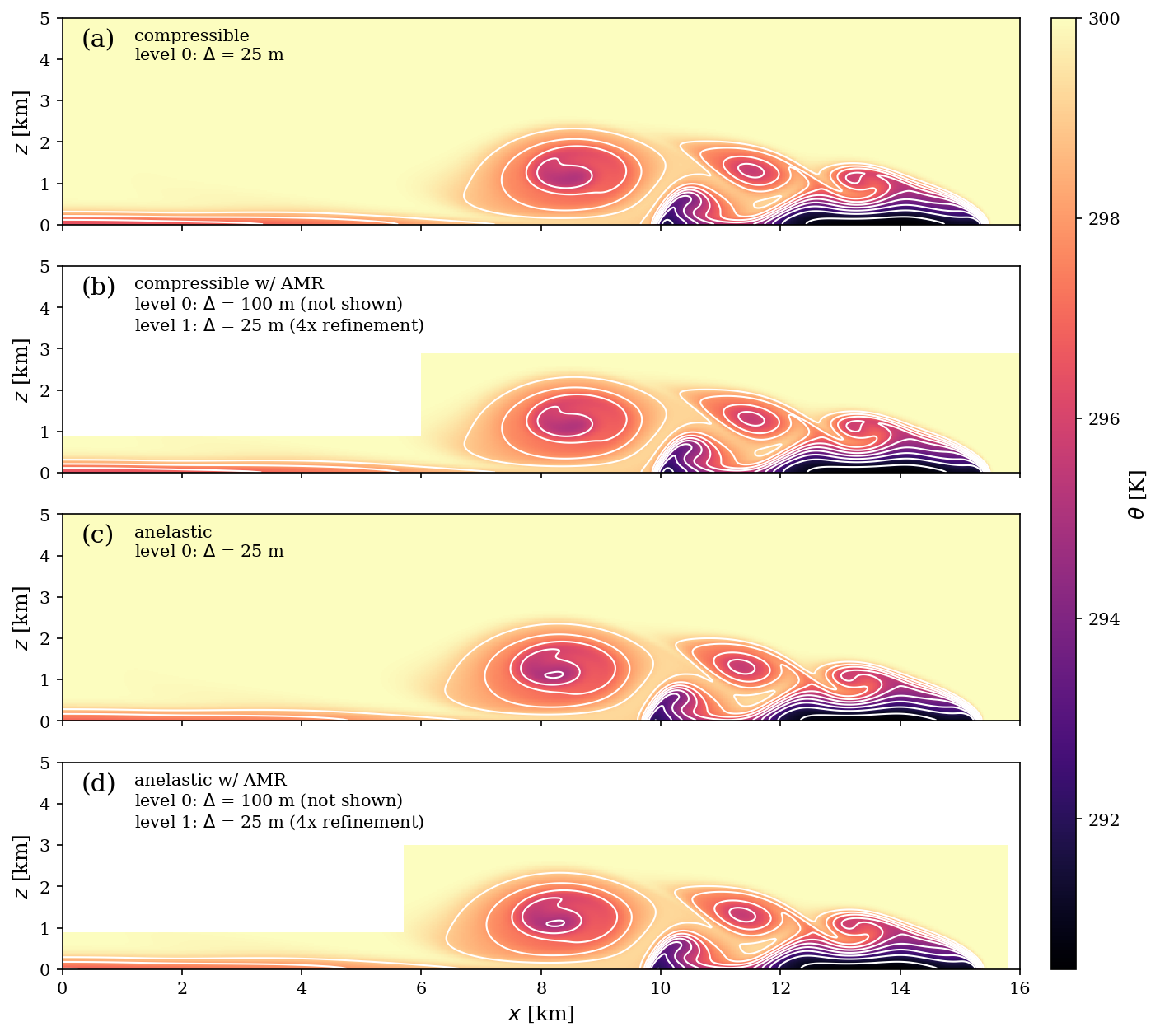}
     \caption{Density current potential temperature fields after 900~s for the reference compressible (fine grid, $\Delta = \Delta x = \Delta z = 25$~m), anelastic ($\Delta = 25$~m), and AMR simulations in both compressible and anelastic simulation modes. Both AMR examples shown here used one level of factor four refinement that locally increased the mesh resolution from $\Delta = 100 \to 25$~m --- only the finest grid level is shown. Temperature contours are plotted at 1~K intervals.}
     \label{fig:dc_comp_anel_ml}
\end{figure}

The coarser grids and anelastic solutions all agree with the reference solution in terms of thermodynamics to within 0.3\% (Fig.~\ref{fig:dc_time_hist}(b)). For both simulation modes, the relative errors tend toward zero with increasing grid resolution. However, the transport of momenta reveal that while similar, the compressible and anelastic equation sets converge to distinct solutions (Fig.~\ref{fig:dc_comp_anel_ml}(a) and (c)), with differences that vary over time and the quantity of interest. These differences are within 5\% for maximum outflow velocity and within 4\% for front propagation speed (Fig.~\ref{fig:dc_time_hist}(d) and (f)) and may be attributed to the production of vorticity from baroclinicity, which differs for the two simulation modes \citep{Kurowski2014_AnelasticCompressibleSimulation}. Finally, we note that the solutions with AMR are indistinguishable from their corresponding fine-grid solution (Fig.~\ref{fig:dc_comp_anel_ml}), with the anelastic and compressible solutions seeing a computational speedup of 2.0x and 2.9x.
\subsubsection{Dry and Moist Bubble rise} \label{subsub:bubble}

Here we compare dry and moist bubble rise simulations in ERF to benchmark data from \citet{bryan2002benchmark,Duarte_2014}. The bubble rise cases rigorously test the buoyancy source term as well as the latent heat associated with phase change from water vapor to cloud water.

For the dry bubble simulation, a base state in HSE is first established for the constant $\theta_{d}=300$ K profile and a surface pressure of $p_b = 1 \times 10^{5}$ Pa. Thermal perturbations, $\Delta \theta_{d}$, are added to generate the warm bubble as follows 
\begin{eqnarray}
\Delta \theta_{d} = 
\begin{cases} 
2 \cos^2\left(\cfrac{\pi r}{2}\right) & r < 1\\
0 & r > 1,
\end{cases}
\end{eqnarray}
where
\begin{align}
    r &= \sqrt{\left(\frac{x - x_c}{x_r} \right)^2 + \left( \frac{z - z_c}{z_r} \right)^2},
\end{align}
with $x_c = 10.0$ km, $z_c = 2.0$ km, and $x_r = z_r = 2.0$ km. The grid spacing is $(100.0, \, 100.0)$ m with a timestep of $0.5$ s and the fluid is treated as inviscid.

\begin{figure}[htpb!]
     \centering
     \includegraphics[width=0.99\textwidth]{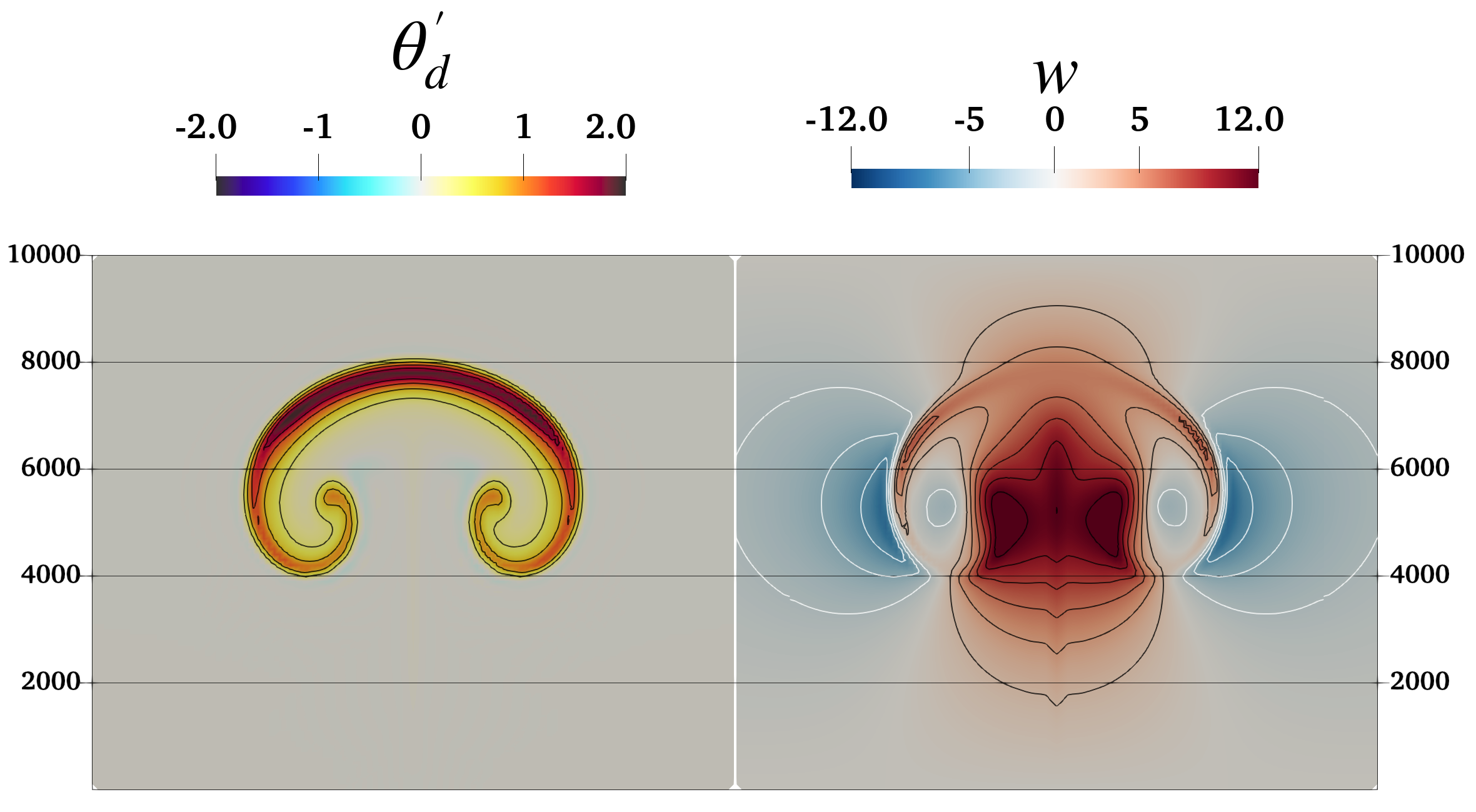}
     \caption{Dry bubble rise with (left) $\theta_{d}^{\prime}$ contoured every $0.4$ K and (right) $w$ velocity contoured every $2$ m/s. Black contours are positive and white contours are negative.
    }
     \label{fig:dry_bubble}
\end{figure}

At constant pressure, the warm perturbations noted above yield negative density perturbations. Consequently, vertical buoyancy forces cause the bubble to rise, along with the formation of characteristic ``rotors'' at the edge of the stretched bubble and strong thermal gradients near the center line. The final state for the pertubational potential temperature and vertical velocity illustrate these features; see Figure~\ref{fig:dry_bubble}. The final rise height of the bubble, $8$ [km], agrees with the results provided in \citet{bryan2002benchmark}.

\begin{figure}[htpb!]
     \centering
     \includegraphics[width=0.99\textwidth]{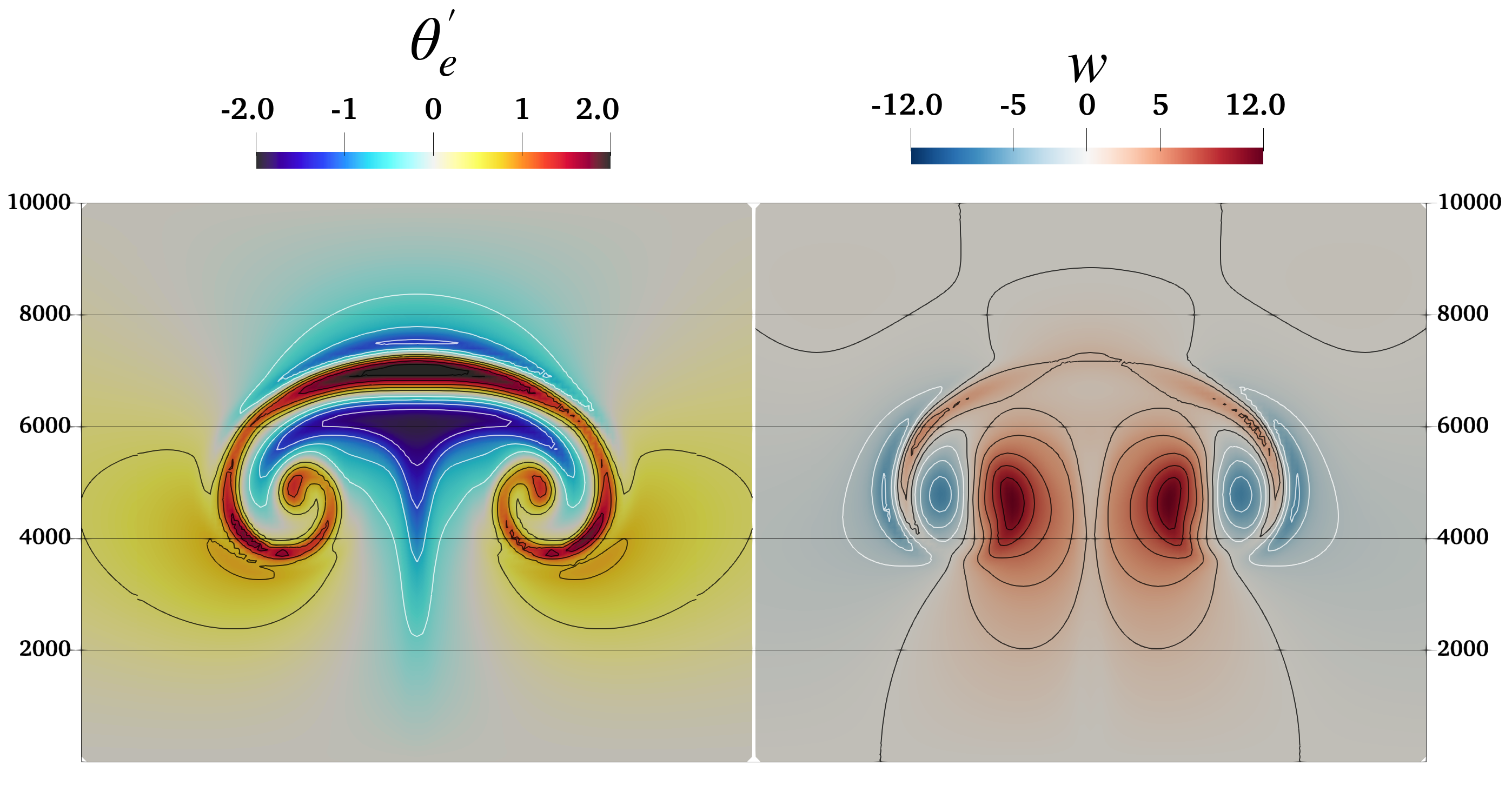}
     \caption{Moist bubble rise with (left) $\theta_{e}^{\prime}$ contoured every $0.4$ K and (right) $w$ velocity contoured every $2$ m/s. Black contours are positive and white contours are negative.}
     \label{fig:moist_bubble}
\end{figure}

For the moist bubble simulation, the initial HSE base state is characterized by a constant equivalent potential temperature, $\theta_e = 320$ K, and constant total moisture content $q_{t0} = q_{v0} + q_{c0} = 0.02$ kg/kg. Establishing the HSE background state for the moist bubble follows the same approach as all other cases, except that it requires embedded Newton-Raphson iterations to satisfy the constraint on equivalent potential temperature  
\begin{eqnarray}
\theta_e &\stackrel{\text{def}}{\equiv}& T_0\Bigg(\frac{p_0}{p_b}\Bigg)^{-\frac{R}{c_p + c_{pl}q_{t0}}}\exp\Bigg[\frac{L_vq_{vs0}}{(c_p + c_{pl}q_{t0})T_0}\Bigg],
\end{eqnarray}
with $q_{vs}$ being the vapor mixing ratio at saturation. Similar to the dry bubble case, buoyancy drives upward motion and rotors form at the bubble edge; see Figure~\ref{fig:moist_bubble}. However, the latent heat due to phase change provides additional thermal sources that increase the fluctuations in equivalent potential temperature and lead to reduced rise height; see Figure~\ref{fig:moist_bubble} (left) versus Figure~\ref{fig:dry_bubble} (left).

Finally, the conservation of total mass and energy (see Eqs. 28 -- 29 in \citet{bryan2002benchmark}) were considered for the bubble rise problems. Total mass is conserved to machine precision but the total energy with moisture is observed to change by $\approx 1\times 10^{-3} \, \%$. We note that the equation set considered by ERF is analogous to ``equation set A'' in \citet{bryan2002benchmark} and the observed conservation properties are also consistent with Figure 7 in \citet{bryan2002benchmark}. 

\subsubsection{Idealized two-dimensional squall line} \label{subsub:squall}

A squall line is a continuous linear formation of thunderstorms or convective cells that may produce severe weather with heavy precipitation. There are a variety of factors that determine the development of a squall line---moisture content, wind shear or squall (a sudden sharp increase in wind near the surface), atmospheric instability, and lifting mechanisms such as cold pool and gust front formation \citep{Rotunno_1988}. In this section we describe an idealized, two-dimensional squall line test case that has been widely studied in the literature \citep{Klemp_1978, Weisman_1984, Gabersek_2011, Tissaoui_2022}. This test case incorporates many features that are significant for a wide range of atmospheric flows---buoyancy, phase change and associated latent heat release, precipitation and cloud formation.

The initial condition consists of an ellipse-shaped, warm, moist bubble in a moist, hydrostatic background with an initial wind shear profile. The potential temperature increment is given by
\begin{eqnarray}\label{eqn:SL_Delta_theta}
\Delta\theta_{d} = 
\begin{cases} 
3 \cos^2\Big(\cfrac{\pi r}{2}\Big) & r<1 \\
0 & r > 1,
\end{cases}
\end{eqnarray}
where 
\begin{equation}
r = \sqrt{\Bigg(\frac{x - x_c}{x_r}\Bigg)^2 + \Bigg(\frac{z - z_c}{z_r}\Bigg)^2},
\end{equation}
with $x_c = 75\times 10^3$ m, $z_c = 2\times 10^3$ m, $x_r = 10\times 10^3$ m, and $z_r = 1.5\times 10^3$ m. The initial profile for the horizontal velocity is given by
\begin{eqnarray}\label{eqn:SL_init_wind}
u(z)~\text{m/s}= 
\begin{cases} 
-12.0 + 0.0048 z & z\le 2500~\text{m}\\
0.0 & z > 2500~\text{m}.
\end{cases}
\end{eqnarray}
Constant diffusivities of $\nu = \alpha_{i} = 200$ m$^2$/s are utilized for all variables; no turbulence model is used. The grid spacing is $(100.0, \, 100.0)$ m with a timestep of $0.25$ s. 

\begin{figure}[H]
\centering
\includegraphics[width=0.99\textwidth]{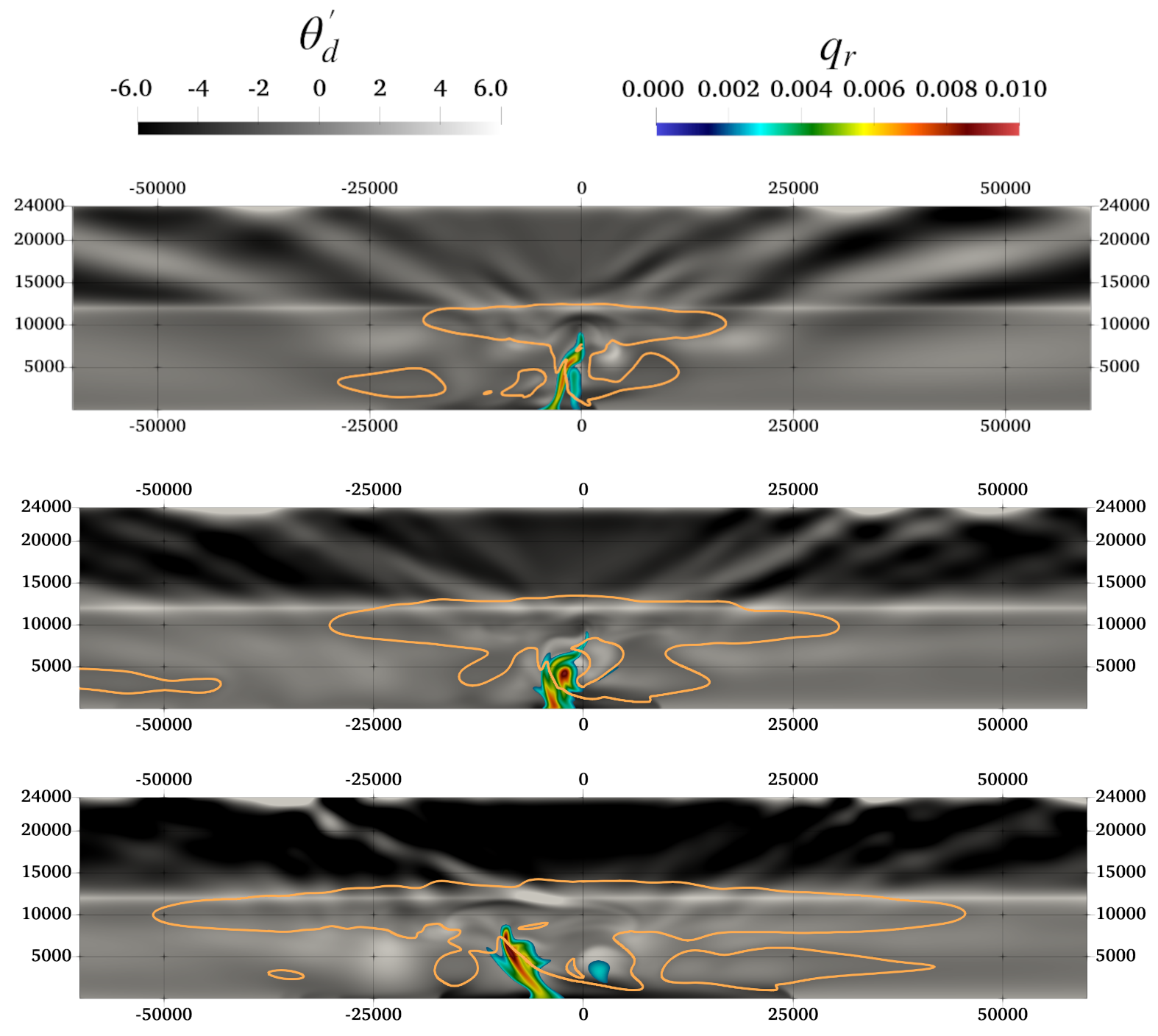}
\caption{The squall line simulation at (top) $3000$ s, (middle) $6000$ s, and (bottom) $9000$ s. The orange contour denotes the cloud $q_{c}$=$10^{-5}$ kg/kg and the perturbation potential temperature is given by $\theta_{d}^{\prime}$ = $\theta_{d}(t)-\theta_{d}(0)$ K. Note, the horizontal $x$-direction has been clipped to $\pm 60$ km to highlight the region of interest around the cloud.}
\label{fig:SL_qc_pot_temp}
\end{figure}

Development of the squall line is illustrated in Figure~\ref{fig:SL_qc_pot_temp}. Initially, the density inside the warm bubble is lower than the background density and buoyant forces drive updrafts. The surface wind shear tilts the buoyant updrafts as they rise up. Furthermore, no condensation occurs initially since the initial condition is characterized by a water vapor mixing ratio that is less than the saturation mixing ratio at all heights. Cloud water formation begins to occur at $t\approx 420$ s due to the vertical transport of water vapor by the warm bubble; the latent heat release from this phase change further contributes to vertical transport. Outline of the cloud, as characterized by the orange $q_c=10^{-5}$ kg/kg isocontour, is also shown in Figure~\ref{fig:SL_qc_pot_temp}. The cloud water begins to form rain through the processes of autoconversion and accretion at $t\approx 660$ s. The formation of cloud water and rain results in increased density (known as precipitation loading), and creates negative buoyancy, resulting in downdrafts.

Once formed, rain begins to fall with a downward terminal velocity and accumulation at the surface begins around $t\approx 1200$ s. As the rain reaches the surface, it absorbs latent heat from the surrounding air and evaporates to form water vapor. This results in a rain-cooled cold front. The evaporation of rain and under-saturated cloud water creates more water vapor that rises and condenses to form clouds. This results in a feedback mechanism that sustains the thunderstorm, with a cold front that propagates to the right. The updrafts are strengthened as they rise into the cloud, which leads to the typical anvil cloud structure.

It can be seen that the maximum height to which the cloud rises is $\approx$ 14 km. During the later stages ($t$=9000 s), a cold pool of rain-cooled air is formed at the bottom due to the evaporation of rain and the associated latent heat absorption from the surrounding air. The cloud evolution, maximum height, and the potential temperature perturbation qualitatively agree with results presented in \cite{Tissaoui_2022}.

\begin{figure}[H]
\centering
\includegraphics[width=0.5\textwidth]{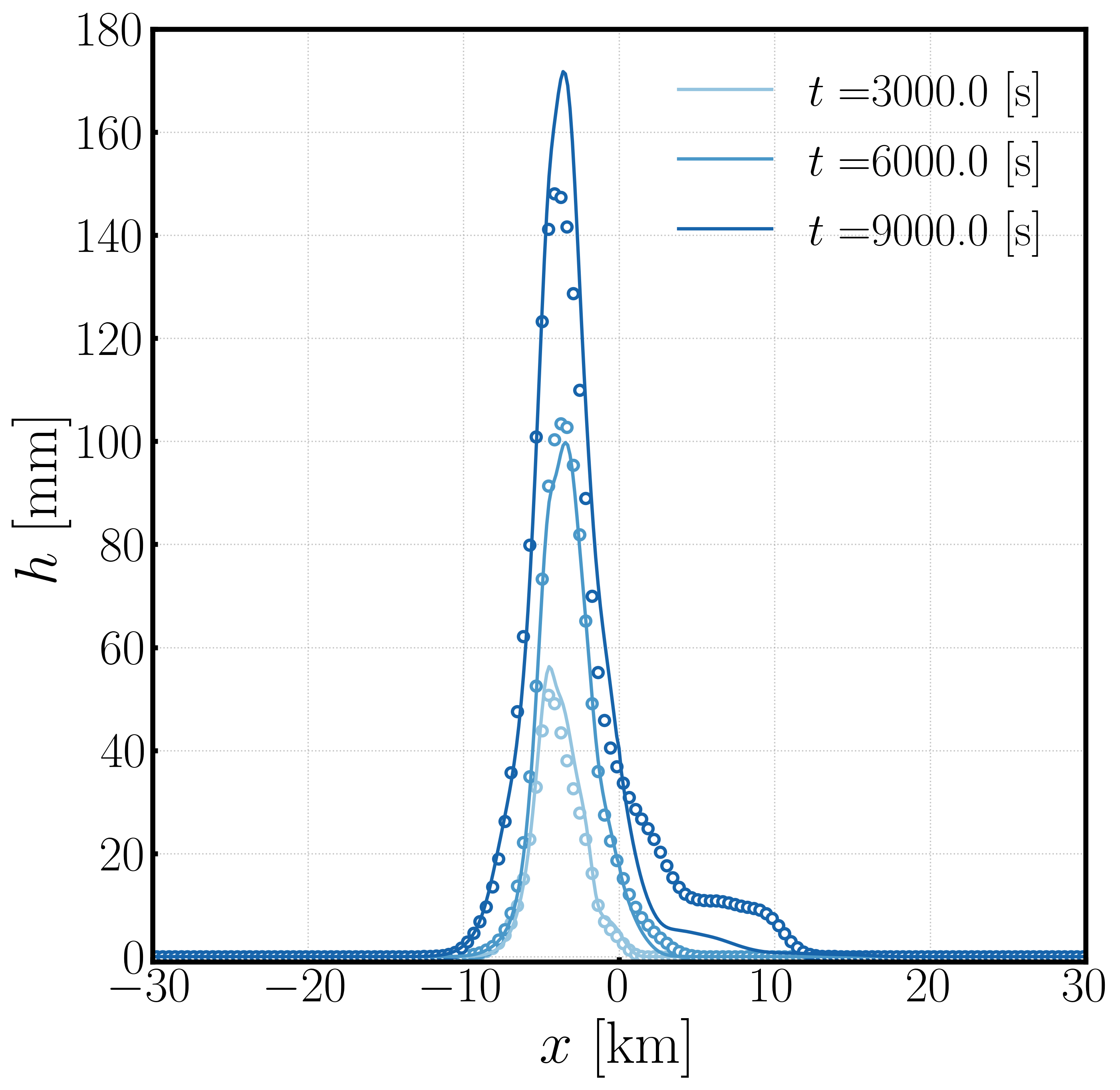}
\caption{The rain accumulation, $h(x)$, obtained from ERF (solid lines) versus WRF (markers) for the squall line case at times $t = 3000, 6000$ and 9000 s.}
\label{fig:SL_rain_accum}
\end{figure}

Accumulation of precipitation at the ground is a quantity of interest in a squall line simulation, and is computed using the values in the first mesh cell above the ground. The total accumulation over time $T$ is obtained by adding the accumulation over each time step $\Delta t$, and is given by
\begin{eqnarray*}
h(x,y) = \sum_{t=0}^T\Bigg(\cfrac{\rho_d q_r w_t \Delta t}{\rho_\text{water}}\Bigg),
\end{eqnarray*}
where $\rho_d$ is the density of dry air, $q_r$ is the mixing ratio of rain water, $w_t$ is the terminal velocity of rain, and $\rho_\text{water}$ is the density of water. 
A comparison of the rain accumulation obtained from ERF and WRF \citep{WRF:Skamarock} is shown in Figure~\ref{fig:SL_rain_accum} at $t=3000$, 6000, and 9000 s. The rainfall is confined to a strip of length 20 km about the center of the domain, which is typical in such simulations. Good quantitative comparison is observed between WRF and ERF. However, it is noted that an exact bit-wise match should not be expected since different coordinate systems and saturation correlations were employed by the two codes; see paragraphs 7 of Section~\ref{sec:intro} and 2 of Section~\ref{subsec:micro}.

%
%
%
%
%
%

\subsubsection{Three-dimensional supercell} \label{subsub:supercell}

The supercell case is a three-dimensional version of the squall line described in Section~\ref{subsub:squall}, and has been recently studied by \citet{Tissaoui_2022,kang2024multiscale}. The initial background state and the wind shear are the same as the squall line test case but the initial warm bubble is three-dimensional and characterized by Eq.~\ref{eqn:SL_Delta_theta}, with
\begin{equation}
r = \sqrt{\Bigg(\frac{x - x_c}{x_r}\Bigg)^2 + \Bigg(\frac{y - y_c}{y_r}\Bigg)^2 + \Bigg(\frac{z - z_c}{z_r}\Bigg)^2},
\end{equation}
with $x_c = 75\times 10^3$ m, $y_c = 50\times 10^3$ m, and $z_c = 2\times 10^3$ m, $x_r = 10\times 10^3$ m, $y_r = 10\times 10^3$ m, and $z_r = 2\times 10^3$ m. The initial horizontal wind velocity profile is the same as in Eq.~\ref{eqn:SL_init_wind}. Constant diffusivities of $\nu = \alpha_{i} = 33.33$ m$^2$/s are employed for all variables; no turbulence model is employed. The grid spacing is $(250.0, \, 250.0, \, 250.0)$ m with a timestep of $0.5$ s.

\begin{figure}[H]
\centering
\begin{subfigure}[b]{0.99\textwidth}
\includegraphics[width=\textwidth]{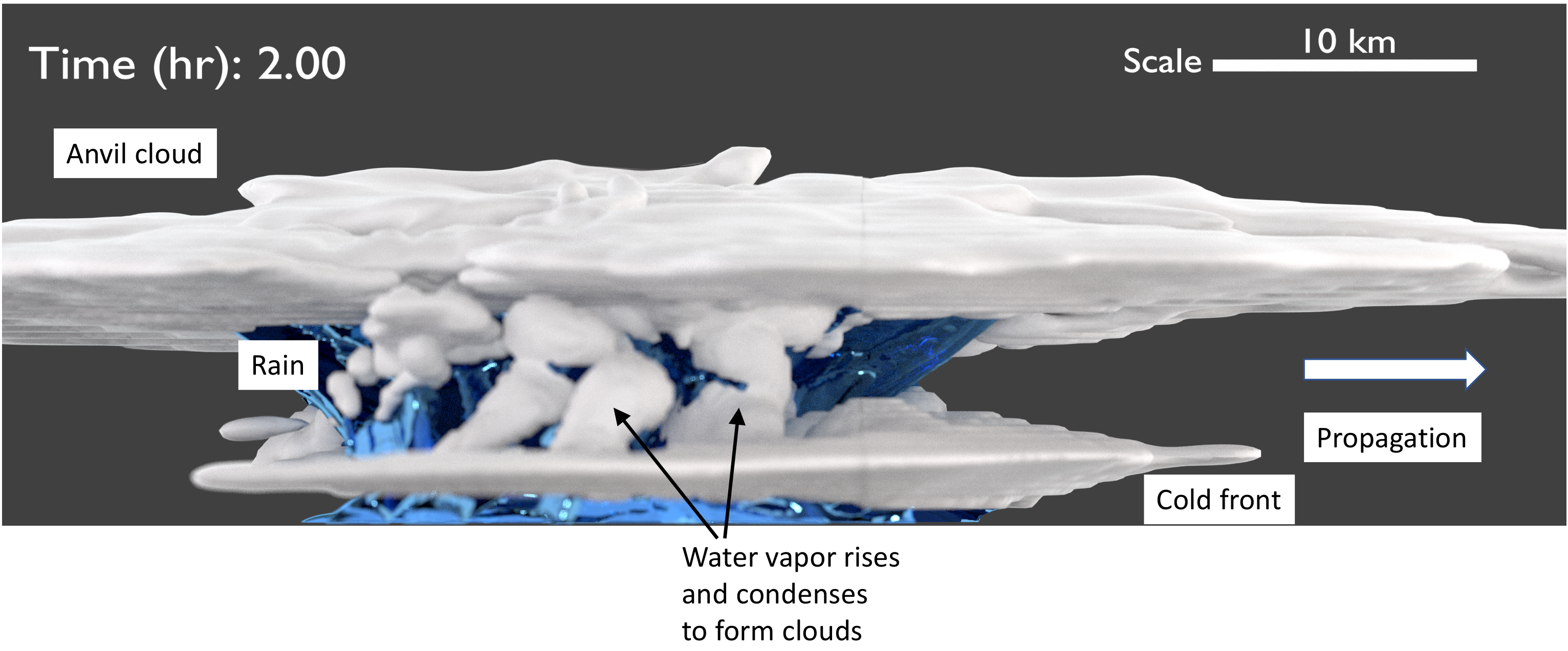}
\end{subfigure}
\caption{Ray trace rendering (in Blender v2.81 \citep{blender}) of the supercell evolution highlighting the main features. Isocontours of cloud water ($q_c$=$10^{-5}$ kg/kg) in white and rain water ($q_r$=$10^{-4}$ kg/kg) in blue are shown.}
\label{fig:SC_ray_trace}
\end{figure}

The evolution of the three-dimensional supercell is qualitatively similar to the two-dimensional squall line; see Section~\ref{subsub:squall}. The initial warm, moist bubble rises upward due to buoyancy, and as the bubble rises, water vapor begins to condense to form cloud water at $t$ $\approx$ 300 s. The cloud water agglomerates to form rain at $t$ $\approx$ 540 s, and precipitation accumulation on the ground begins at $t$ $\approx$ 800 s. The under-saturated cloud water and the falling rain evaporate leading to the formation of more water vapor that rises and condenses, creating a feedback loop that intensifies the supercell. Figure~\ref{fig:SC_ray_trace} shows a ray-trace rendering of the supercell at $t$=7200 s, highlighting the the main features of the supercell evolution. The isocontour of cloud water mixing ratio ($q_c=10^{-5}$ kg/kg) in white, rain water mixing ratio ($q_r=10^{-4}$ kg/kg) in blue, the anvil cloud structure, rain-cooled cold front, and the storm propagation are shown.

\begin{figure}[H]
\centering
\includegraphics[width=0.99\textwidth]{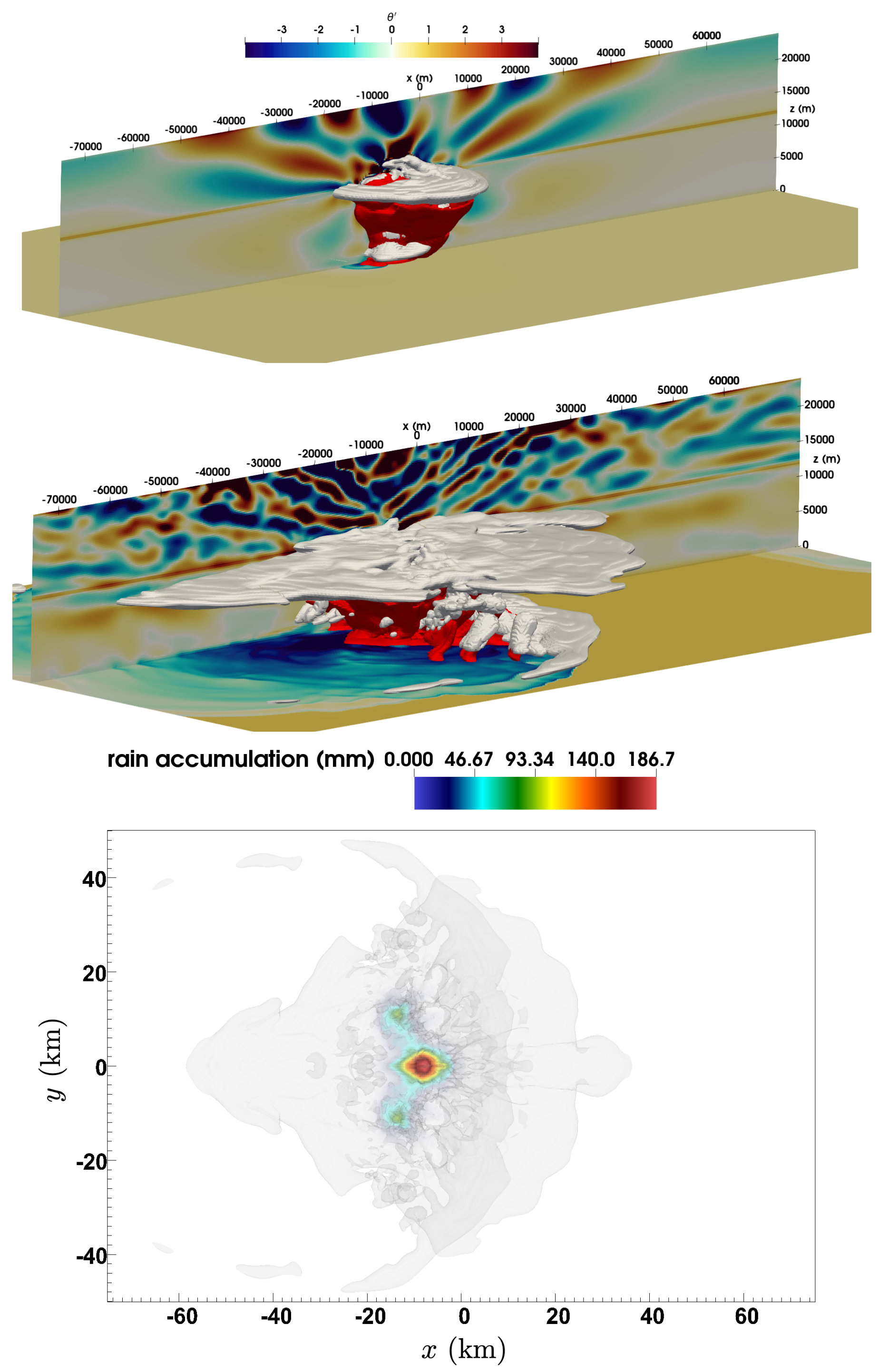}
\caption{Evolution of the three dimensional supercell: Isocontour of cloud water mixing ratio ($q_c$ = $10^{-5}$ kg/kg) in white, rain water mixing ratio ($q_r$ = $10^{-4}$ kg/kg) in red, and contours of potential temperature perturbation ($y$=0 and $z$=0 slices) ($\theta_{d}^{\prime}$ = $\theta_{d}(t)-\theta_{d}(0)$ K) at $t$=1800 s (top), and 7200 s (middle), and the total rain accumulation on the ground in mm  at $t=7200$ s (bottom).} 
\label{fig:SC_qc_qr_pot_temp}
\end{figure}

Figure~\ref{fig:SC_qc_qr_pot_temp} shows the evolution of $q_c$, $q_r$, and $\theta_{d}^{\prime}$ ($y$=0 and $z$=0 slices) at  $t$=1800 and 7200 s. It can be seen from the cloud water isocontour that the cold front evolves into a characteristic bow-shaped structure and propagates to the right. The maximum height to which the cloud rises is $\approx$ 14 km. The $z$=0 slice of the potential temperature perturbation shows the formation of a region near the ground that is evaporatively cooled by rain. The $q_{r}$ isocontour shows that the rain formation is confined to the central region of the supercell, spanning $\approx$ 20 km and 40 km in the $x$- and $y$-directions, respectively. Precipitation on the ground is further confined, with the heaviest accumulation observed within a region of 10 km and 5 km in the $x$- and $y$-directions, respectively.

%
%

\subsection{Microscale ABL} \label{subsec:abl}
Having considered representative mesoscale phenomena in Section~\ref{subsec:atmosphere}, we turn our attention to the microscale. Here we assess the ability of ERF to resolve realistic features of the turbulent ABL under varying atmospheric stability conditions, with and without moisture. We simulate canonical neutral and stable dry ABLs before turning to a convective ABL capped by shallow cumulus clouds. The latter case includes more realistic atmospheric forcings derived from field data. For all three ABL studies, we present our results alongside code intercomparisons found in literature. 

\subsubsection{Conventionally Neutral Atmospheric Boundary Layer} \label{subsub:neutralabl}

Evaluations of turbulence-resolving LES typically begin with the idealized, conventionally neutral boundary layer, also known as an inversion-capped or conditionally neutral ABL. This ABL is characterized by uniform mean potential temperature with overlying layers of stably stratified air. The capping inversion is a thin, strongly stable layer with a large temperature gradient, which prevents the growth of the ABL over time through the buoyant destruction of turbulence. Here, we evaluate an idealized neutral ABL that corresponds to realistic conditions observed at the U.S. Department of Energy's Scaled Wind Farm Technology facility \citep{Mirocha2018_LESsensitivities}.

We specifically base our model validation effort on the high-resolution cases from \citet{Mirocha2018_LESsensitivities}. Following this benchmark study, we include two model configurations: a higher-order upwind scheme with $5^{\rm th}$ ($3^{\rm rd}$)-order horizontal (vertical) advection operators, which mimics a typical WRF model configuration; and a $2^{\rm nd}$-order central-differencing scheme that follows an established LES code, the Simulator for Wind Farm Applications \citep{Churchfield2012_NumericalStudyEffects}. The higher-order upwind approach utilizes a grid with horizontal (vertical) spacing $\Delta x = 15$~m ($\Delta z = 5$~m)---i.e., an aspect ratio of 3, which follows WRF best practices \citep{Mirocha2010_ImplementationNonlinearSubfilter,Kirkil2012_ImplementationEvaluationDynamic}. The $2^{\rm nd}$-order central approach utilizes a uniform grid---standard practice for a variety of LES applications---with $\Delta x = \Delta z = 7.5$~m. We initialize the solution from an idealized sounding: surface pressure of 1000~hPa; constant wind speed, equal to the geostrophic wind vector ($U_g$, $V_g$) = (6.5, 0) m/s; constant potential temperature of 300~K up to base of the capping inversion at 500~m; a capping inversion between 500 and 650~m with strength of 10~K/km; and an overlying weakly stable layer with lapse rate of 3~K/m. The solution is advanced with $\Delta t=0.2$ and $0.1$~s for the high and low-order schemes, respectively. These timestep sizes corresponded to an acoustic CFL number of over 4 but an advective CFL of less than 0.1. To address the propagation of acoustic waves, we use 10 acoustic substeps per time step. A preliminary study (not shown) found that, for canonical conditions such as those under consideration here, larger timesteps with 10 or more accompanying substeps could be used with no ill effect. To encourage the development of turbulence, we follow \citet{Churchfield2012_NumericalStudyEffects} and add deterministic, divergence-free, sinusoidal perturbations (with amplitude of $0.1U_g$ and 12 wave periods in the horizontal directions) to the lowest 500~m of the computational domain. While these ABL cases represent only a small subset of practical configurations of ERF, a comprehensive model sensitivity analysis is beyond the scope of the current paper.

\begin{figure}[H]
\centering
\includegraphics[width=0.9\textwidth]{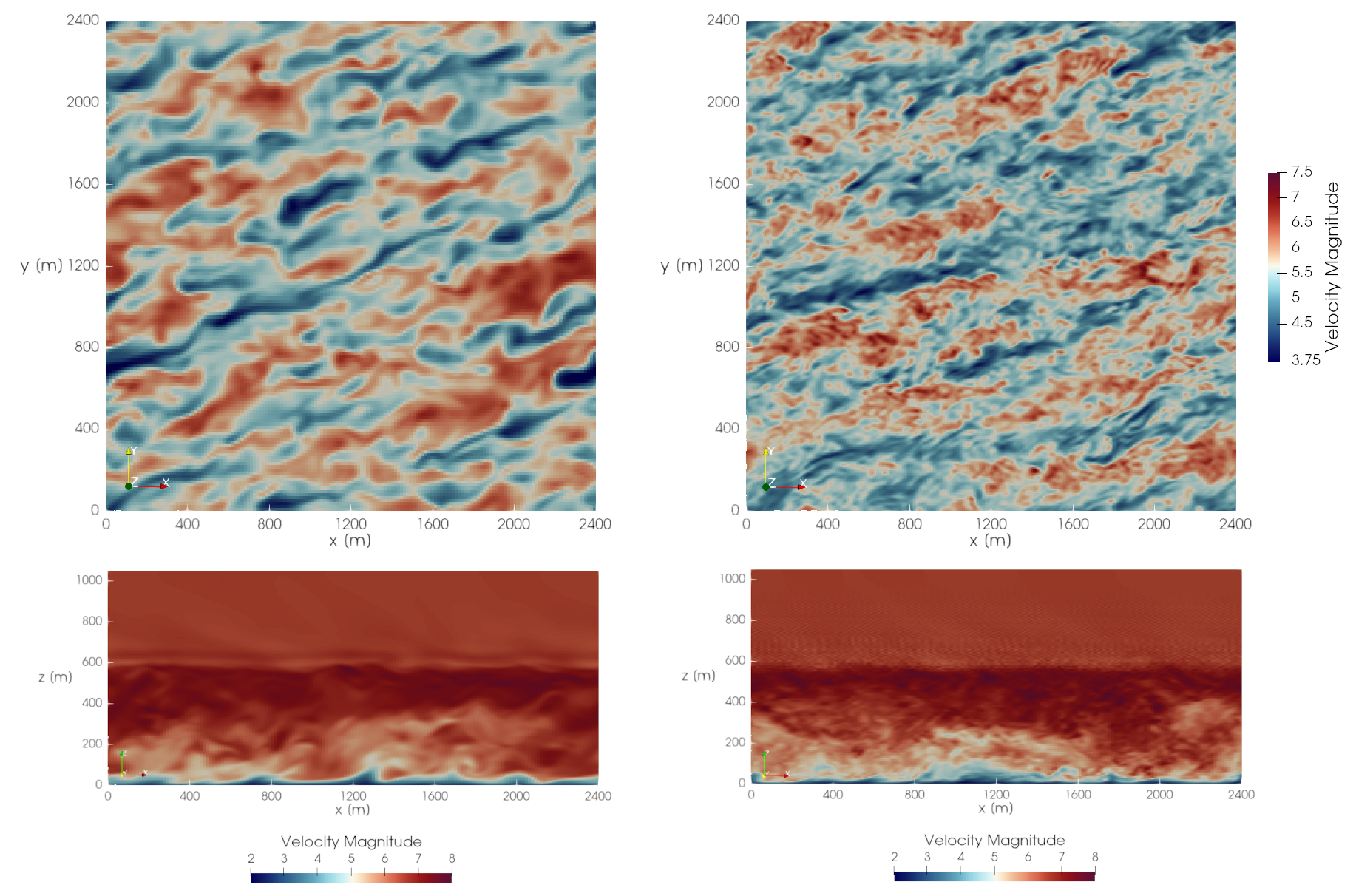}
\caption{Contours of velocity magnitude from horizontal sampling surfaces at $z$=80~m AGL (top) and vertical sampling surfaces aligned with the geostrophic wind (bottom); the left panels correspond to a WRF-like advection scheme with grid aspect ratio $\Delta x/\Delta z$=15~m / 5~m = 3 whereas the right panels correspond to a $2^{\rm nd}$-order scheme with uniform 7.5~m grid spacing. Each solution snapshot corresponds to the center of the analysis window.\label{fig:neutral_abl_fields}}
\end{figure}

As in the benchmark study of \citet{Mirocha2018_LESsensitivities}, we advance the solution in time until a local maximum in the 10-min, planar-averaged wind speed at 80~m above ground level (AGL) is reached. The analysis period starts and ends 1~hr before and after the time corresponding to the wind-speed maximum. This reference time is at 14.2~hr and 14.7~hr for higher-order upwind and $2^{\rm nd}$-order central difference simulations, respectively \eliot{Update the reference times afte rerun}. Horizontal and vertical slices through the solution field (Figure~\ref{fig:neutral_abl_fields}) show the elongated flow structures associated with a shear-driven ABL. Vertical slices illustrate the effectiveness of the capping inversion at limiting the growth of the ABL. With central differencing, numerical noise is seen in the free atmosphere, emanating from the inversion layer, but is effectively eliminated through the Rayleigh damping layer; see Figure~\ref{fig:neutral_abl_fields}d.

\begin{figure}[H]
     \centering
     \includegraphics[width=0.95\textwidth]{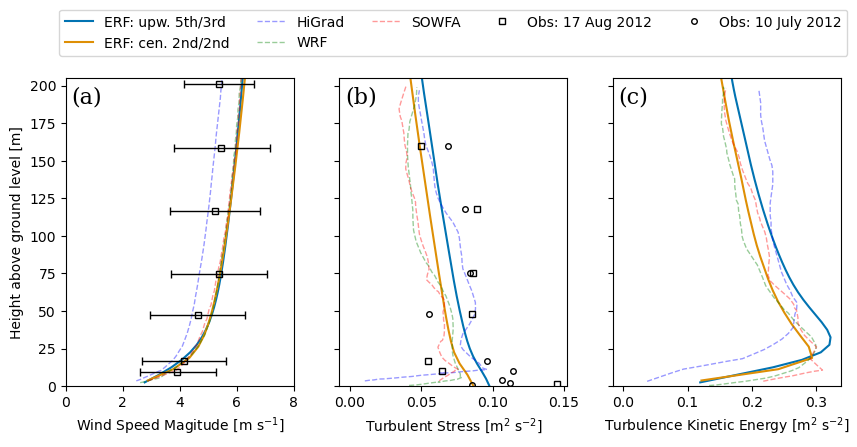}
     \caption{Planar-averaged profiles for the neutral ABL case in comparison with simulation and observational data from \citet{Mirocha2018_LESsensitivities}. \label{fig:neutral_abl_profiles}} 
\end{figure}

\begin{figure}[H]
     \centering
     \includegraphics[width=0.6\textwidth]{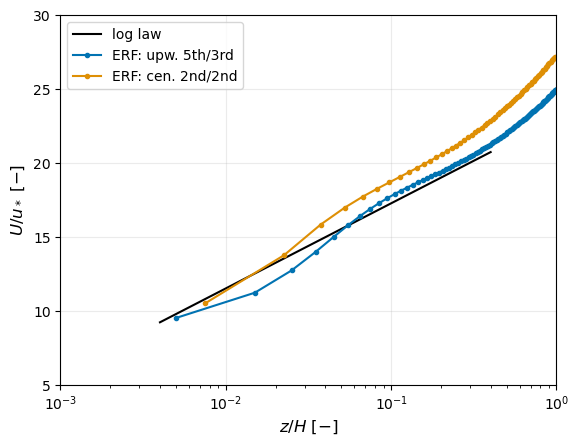}
     \caption{Normalized mean wind profiles for the neutral ABL case in comparison with the log law and simulation data from \citet{Mirocha2018_LESsensitivities}. \label{fig:neutral_abl_loglaw}}
\end{figure}

The simulated mean wind profiles from two ERF model configurations are indistinguishable from two out of three reference simulations (Figure~\ref{fig:neutral_abl_profiles}a). The time- and planar-averaged profiles, based on a time average over the two-hour analysis window, are within one standard deviation of the measured 10-minute wind speeds (indicated by the error bars) at all measurement heights.
In relation to the theoretical logarithmic wind profile (Figure~\ref{fig:neutral_abl_loglaw}), a characteristic mismatch and overshoot is observed \citep[e.g., as in][]{Andren1994_LESNeutrallyStratified,Sullivan1994_SGSModelLES,Kosovic1997_SGSModellingLES,Brasseur2010_DesigningLES,Mirocha2010_ImplementationNonlinearSubfilter} and, as various authors have noted, improvements can be realized through dynamic, nonlinear SGS modeling \citep{Kosovic1997_SGSModellingLES,chow_explicit_2005,Kirkil2012_ImplementationEvaluationDynamic,Lu2014_DevelopmentDynamicNonlinear} and judicious selection of model parameters \citep{Brasseur2010_DesigningLES}.

\eliot{TODO: update validation repo}

The turbulent velocity fields shown in Figure~\ref{fig:neutral_abl_fields} may be quantified with single-point statistics. Results from both ERF models show excellent agreement with comparable models (Figure~\ref{fig:neutral_abl_profiles}b,c) in terms of turbulent stress profiles 
$u_*^2 = (\langle u^\prime w^\prime \rangle)^2 + (\langle v^\prime w^\prime \rangle)^2$
(based on both resolved and SGS stresses) and TKE profiles 
$k = \frac{1}{2}\left[(\langle u^\prime u^\prime \rangle)^2 + (\langle v^\prime v^\prime \rangle)^2 + (\langle w^\prime w^\prime \rangle)^2\right]$
(based on resolved variances only). The ERF profiles are smoother than the reference simulation results due to the use of planar averaging. Reasonable agreement is seen between modeled and measured profiles for the turbulence stress (Figure~\ref{fig:neutral_abl_profiles}b) and TKE (Figure~\ref{fig:neutral_abl_profiles}c); field measurements have been excluded from the TKE comparison for reasons discussed in \citet{Mirocha2018_LESsensitivities}. 

\begin{figure}[H]
     \centering
     \includegraphics[width=\textwidth]{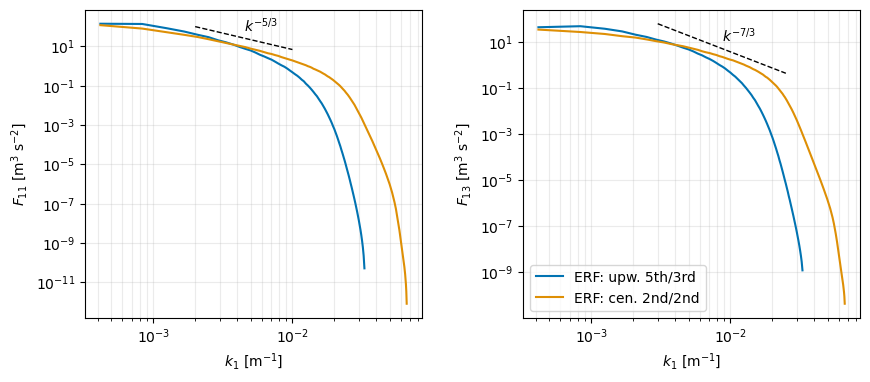}
     \caption{Spectra of the streamwise velocity (left) and cospectra of the $\langle u'w'\rangle$ component of turbulent stress (right) for the neutral ABL case, based on planar data from 80~m AGL, sampled over two hours. \label{fig:neutral_abl_spectra}}
\end{figure}

Frequency domain analysis provides additional insights into the preceding results. Spectra and co-spectra are computed from line samples in the $x$ direction (aligned with the geostrophic wind vector) and ensemble averaged over all points in the lateral $y$ direction, at one-minute intervals over the two-hour analysis period. The additional small-scale turbulence resolution afforded by the central-differencing scheme (see Figure~\ref{fig:neutral_abl_fields} panel b vs a) appears as a much wider frequency range in the power spectral densities (Figure~\ref{fig:neutral_abl_spectra}). The low-order (even-ordered) scheme therefore captures more of the inertial subrange, the portion of the spectrum with a -5/3 slope. The higher TKE predicted by the higher-order upwind scheme (Figure~\ref{fig:neutral_abl_profiles}b) is therefore attributed to the higher energy content of the larger coherent turbulent structures, corresponding to the lower wavenumber portion of the spectra. While both simulations follow Kolmogorov's -5/3 law in the $x$-velocity spectra, albeit over different ranges, only the higher-order upwind result begins to capture the -7/3 power law associated with the $x$ and $z$ velocity spectra.

\subsubsection{Stable Atmospheric Boundary Layer} \label{subsub:stableabl}

The Global Energy and Water Cycle Experiment (GEWEX) Atmospheric Boundary Layer Study (GABLS) is a test case for validating numerical simulations of stratified atmospheric boundary layers with single-column and LES \citep{cuxart_single-column_2006,beare_intercomparison_2006,holtslag_stable_2013}. This benchmark has highlighted the challenges of simulating a moderately stable ABL, which is expected to have continuous---rather than intermittent---turbulence. This turbulence persists in the presence of weak to moderate surface cooling that causes buoyant destruction of turbulence and is coupled with, and offset by, a moderate to strong geostrophic wind that tends to produce more mechanical turbulence \citep{KosovicCurry2000}. These physical mechanisms need to be properly resolved, which requires sufficient grid resolution and has considerable cost. 

Our model configuration follows \citet{Beare2006_GABLS1}. The initial conditions are an idealized sounding described by: surface pressure of 1008~hPa; potential temperature of 265~K from the ground up to 100~m; a strongly stable layer aloft with lapse rate of 10~K/km; and a constant wind field, equal to the geostrophic wind $(U_g,V_g) = (8,0)$~m/s. All cases use a WRF-like $5^{\rm th}$ ($3^{\rm rd}$)-order horizontal (vertical) advection scheme.

\begin{figure}[H]
    \centering
     \includegraphics[width=0.9\textwidth]{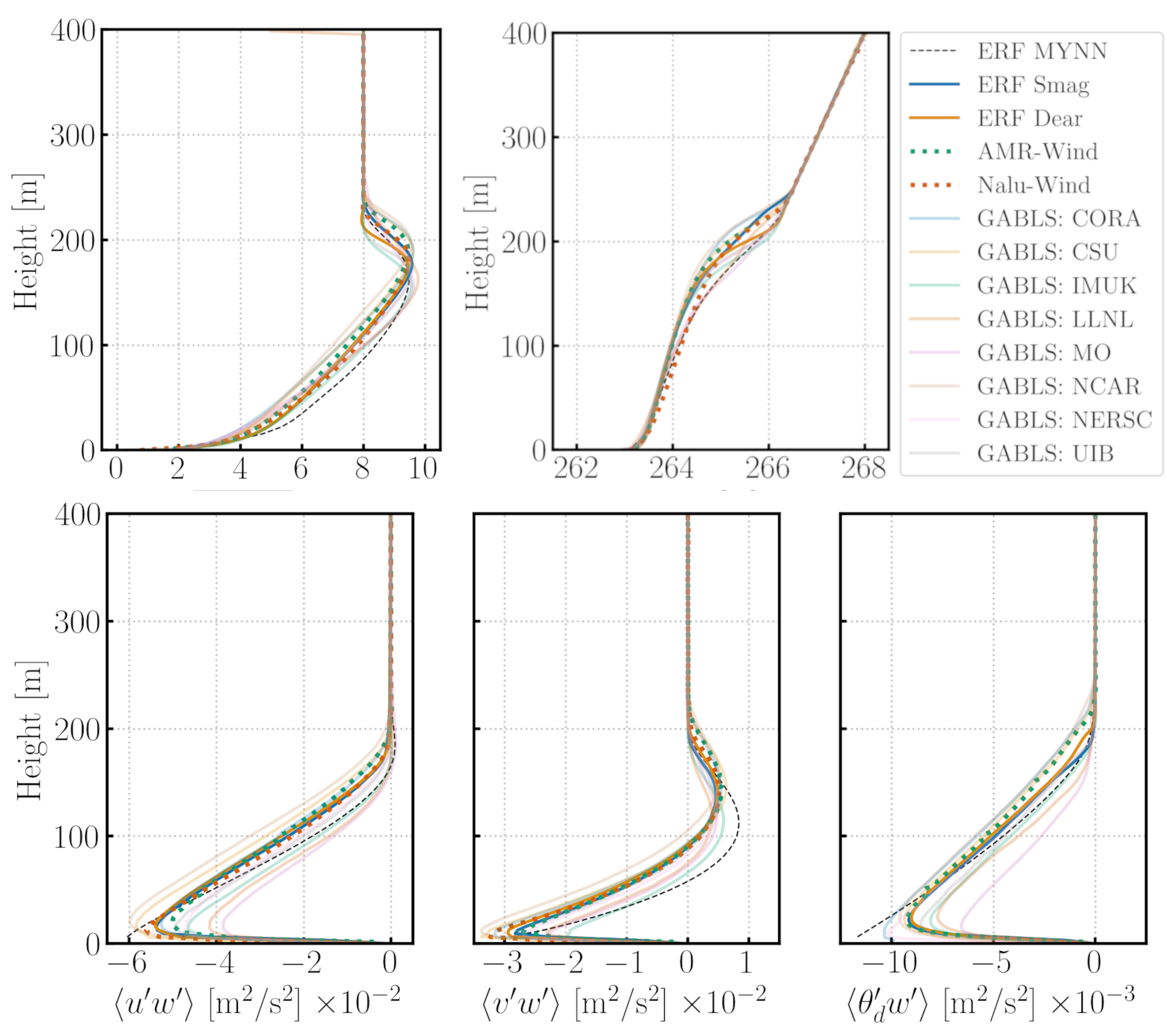}
    \caption{Planar-averaged profiles of (top left) horizontal wind speed, (top right) dry potential temperature, (bottom) kinematic momentum fluxes in $x$ and $y$, and kinematic heat flux for the GABLS1 stable ABL simulation.\label{fig:gabls1_profiles}}
\end{figure}

For LES cases, the Deardorff and Smagorinsky models are used for turbulence closure. 
The solution in the LES cases is advanced for roughly 9 hours with $\Delta t=0.05$ s, 8 substeps per timestep, and a uniform grid size of 3.125 m. The SCM uses $\Delta t=1$ s, 6 substeps per timestep, and a uniform grid size of 6.25 m. For the LES cases, this corresponds to acoustic and advective CFLs of over 5 and 0.13, respectively. We present ERF results alongside previous results from the GABLS intercomparison, along with publicly available results for two additional codes within the DOE ExaWind software suite, AMR-Wind and Nalu-Wind (Figure~\ref{fig:gabls1_profiles}). 

Results from ERF show excellent agreement with previous LES simulations (Figure~\ref{fig:gabls1_profiles}). The ERF SCM result uses the MYNN Level-2.5 PBL scheme and compares well with previous GABLS1 SCM simulations in \citet[][]{cuxart_single-column_2006}. Notable differences are the positive wind-speed bias (Figure~\ref{fig:gabls1_profiles}a), attributed to excessive mixing (Figure~\ref{fig:gabls1_profiles}c--e) typical of SCM simulations. All LES profiles lie within the range of previously simulated values. Furthermore, the ERF LES results tend to agree most closely with the recent Exawind simulations. Sensitivity to the choice of SGS model is most clearly seen in the mean velocity profile above the jet nose (around 175~m AGL, Figure~\ref{fig:gabls1_profiles}a) and at the base of the temperature inversion (between 200--250~m AGL, Figure~\ref{fig:gabls1_profiles}b). These differences in the ERF LES mean profiles can be attributed to differences in the exchange of heat at the top of the ABL (Figure~\ref{fig:gabls1_profiles}e), for which the ABL with Smagorinsky closure shows excessive dissipation. In the future, this may be remedied with a stability correction to the eddy diffusivities based on Brunt--V\"ais\"al\"a frequency \citep{WRF:Skamarock}.

\subsubsection{Shallow cumulus convection} \label{subsub:bomex}

Shallow cumulus clouds play a significant role in determining the thermodynamic structure of the atmosphere; they influence large-scale circulation in the tropics and mid-latitudes through changes in the air--sea thermodynamic and momentum fluxes, lower-tropospheric thermodynamic and wind profiles, capping inversion depth, and radiative transfer between the surface and free troposphere \citep{Siebesma_2003}. The Barbados Oceanographic and Meteorological Experiment (BOMEX) was a 5-day experiment to determine the rate of transfer of water vapor, heat, and momentum from the tropical ocean to the atmosphere, during which trade wind shallow cumulus convection occurred without well-developed storms \citep{HOLLAND_1973, Nitta_1974}. The study has since become a crucial validation test case for atmospheric flow solvers and has been widely studied in the literature \citep{Siebesma_1995,Jiang_2000,Siebesma_2003,Sridhar_2022, munoz-esparza_fasteddy_2022}. Conditions for the BOMEX case are approximated by incorporating forcing terms that describe the interaction between small-scale cloud formation processes and the large-scale atmosphere \citep{Siebesma_2003}.

The 3$^\text{rd}$-order slope limited scheme with a monotonicity preserving switch is used for the advection operator and the Smagorinsky model with $C_{s} = 0.17$ is employed. The timestep is set to 0.3 s and the grid spacings are $(100.0, \, 100.0, \, 40.0)$ m.

\begin{figure}[H]
\centering
\includegraphics[width=0.99\textwidth]{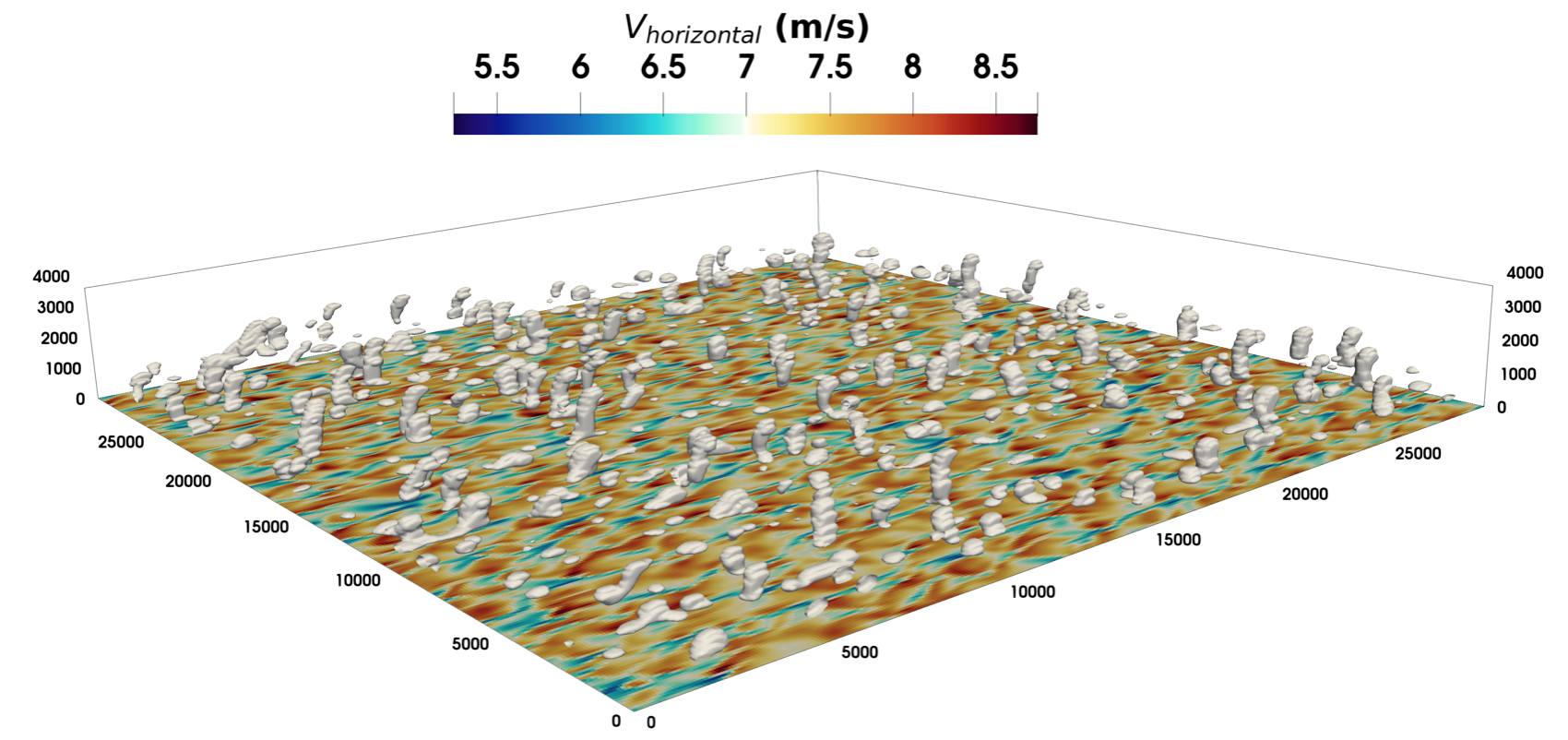}
\caption{Isocontour of cloud water mixing ratio ($q_c$ = $6\times10^{-6}$ kg/kg) and the turbulent velocity field for the  BOMEX simulation on a 30 km $\times$ 30 km $\times$ 3.6 km domain with a resolution of $(\Delta x, \Delta y, \Delta z) = (100,100,40)$ m, at $t=1.6$ hrs.}
\label{fig:bomex_3d}
\end{figure}

The initial condition is under-saturated ($q_v < q_{vs}$) with no $q_{c}$ present. As the simulation proceeds, moisture and thermal energy are introduced into the system through the air--sea interaction at the bottom boundary and prescribed large-scale tendencies, leading to the formation of a convective boundary layer. Vertical transport of moisture due to resolved turbulence yields patches of saturation at $z\approx 500$ m. The thermal energy released from water vapor condensing to form clouds creates further buoyant transport by which the clouds are lifted and advected by the flow-field. The trade inversion layer acts as a cap that limits the vertical growth of clouds and results in the formation of shallow cumulus clouds with a maximum cloud height of $\approx$~2 km. To clearly illustrate the sparse cloud formation present in this case, we employ a larger domain than reported in Table~\ref{tab:BigObnoxiousThing} with extents ($L_x$, $L_y$, $L_z$) = (30, 30, 4) km but the same grid resolution; see Figure~\ref{fig:bomex_3d}. There is no significant precipitation ($\approx$ 0.2 mm/day in the measurements by \citet{HOLLAND_1973}) since the air cannot rise to levels where deep convection can occur, hence the use of a microphysics model without precipitation is justified.

\begin{figure}[H]
     \centering
     \includegraphics[width=0.95\textwidth]{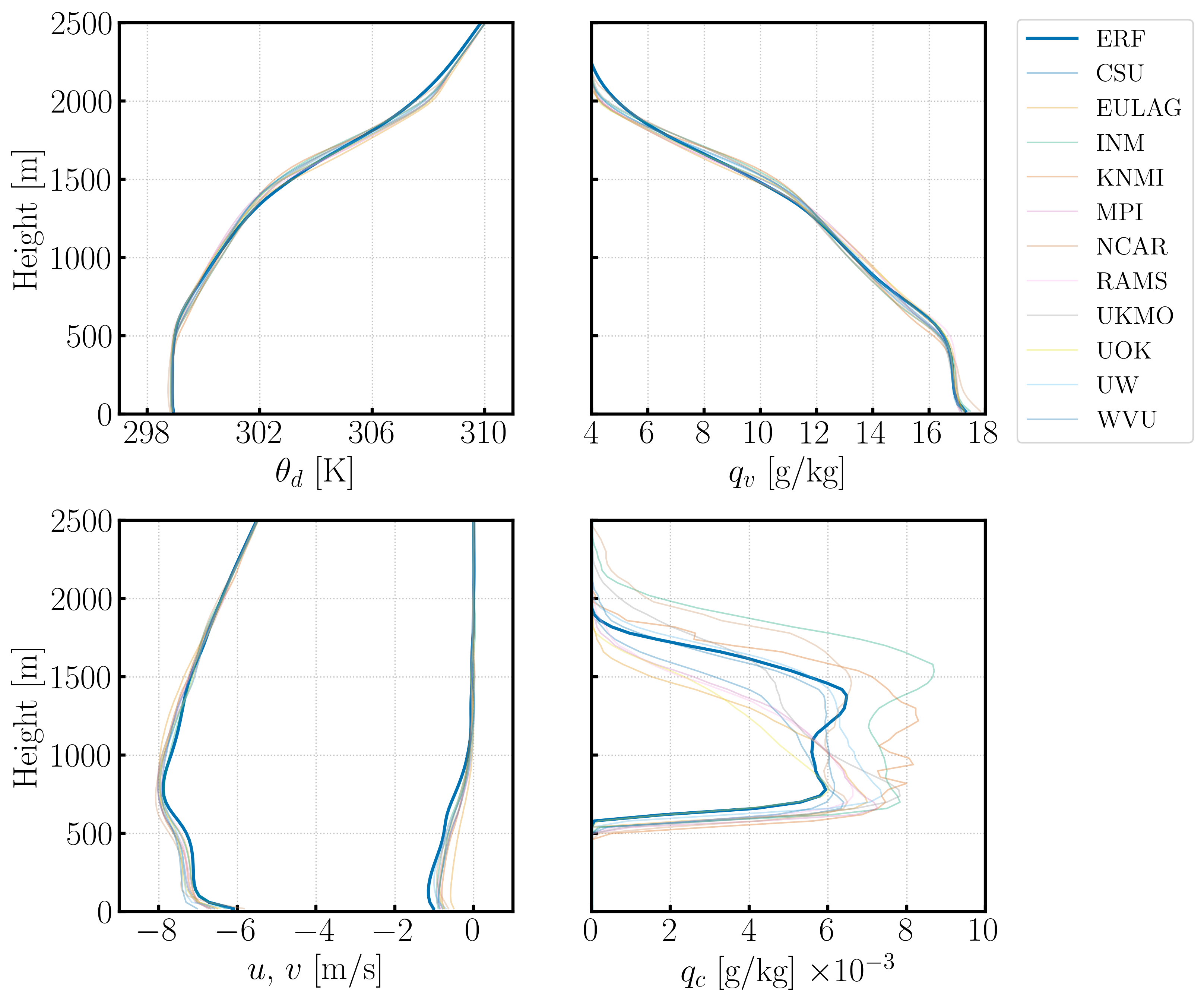}
     \caption{State variables spatially-averaged in the $xy$-plane, and time-averaged in the last hour for the BOMEX case: (Row wise) Potential temperature, vapor mixing ratio, horizontal wind velocities, and cloud water mixing ratio.}
     \label{fig:BOMEX_mean_profs}
\end{figure}

\begin{figure}[H]
     \centering
     \includegraphics[width=0.95\textwidth]{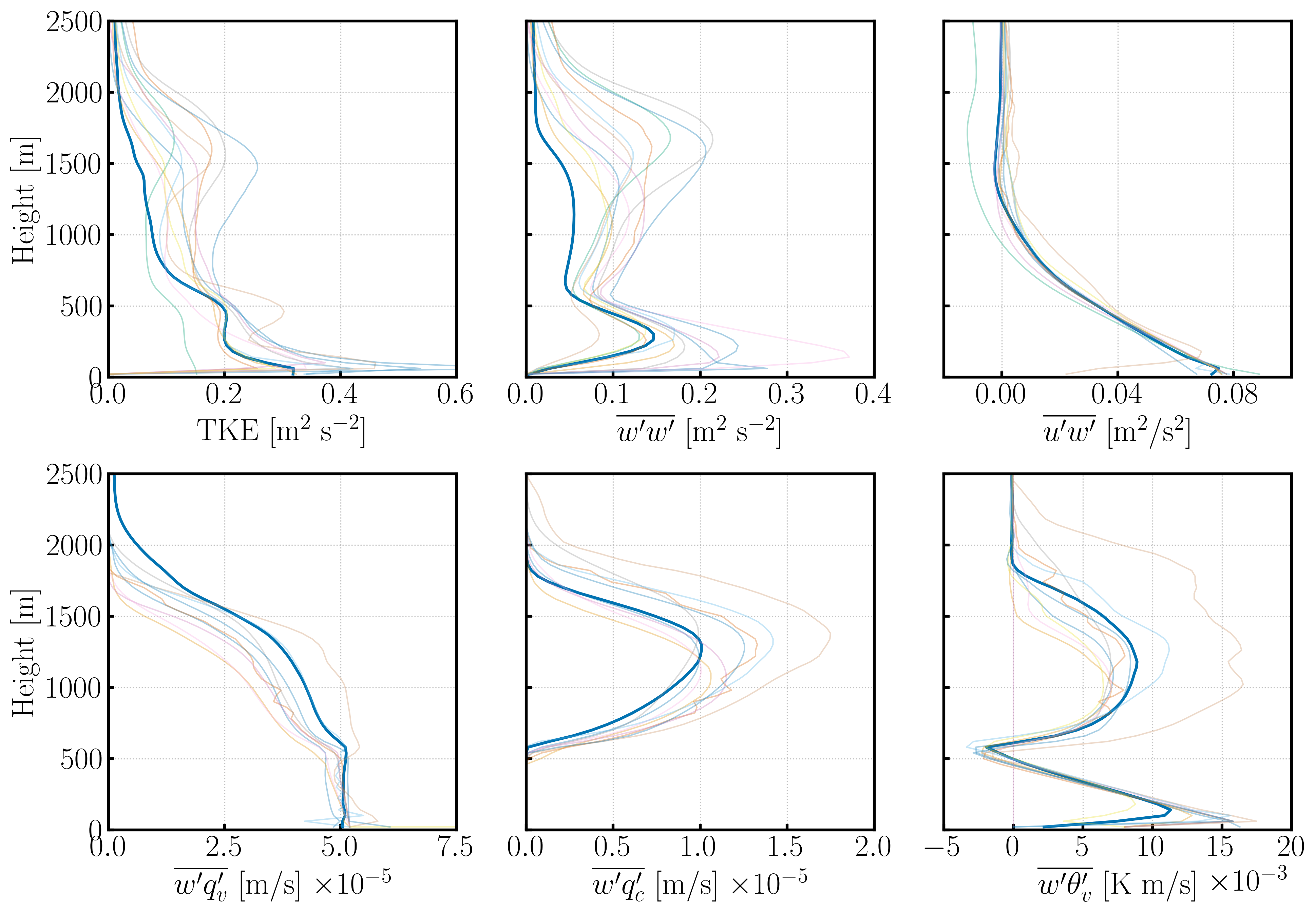}
     \caption{Turbulence quantities spatially-averaged in the $xy$-plane, and time-averaged in the last 3 hours for the BOMEX case: (Row wise) turbulent kinetic energy, turbulent fluxes: $\overline{w'w'}$, $\overline{u'w'}$, $\overline{w'q_v'}$, $\overline{w'q_l'}$, and $\overline{w'\theta_v'}$. Line-styles match the legend provided in Figure~\ref{fig:BOMEX_mean_profs}.}
     \label{fig:BOMEX_turb_fluxes}
\end{figure}

Figure~\ref{fig:BOMEX_mean_profs} shows the spatially ($xy$ plane) and temporally (last 1 hr) averaged profiles of the prognostic variables as a function of the vertical height and the comparison with results from several other codes. Comparison of these profiles with the initial profiles illustrates the changes in the atmosphere due to shallow cumulus convection dynamics. The potential temperature profile shows an increase of $\approx 0.3$ K near the air--sea surface, which is indicative of the heating due to the sensible heat flux. Since saturation only occurs for $z > 500$ m, $q_c$ only has non-zero values above this height. Changes in the velocity fields near the surface are due to the turbulent boundary layer evolution. While evaporation of sea water does introduce moisture into the atmosphere, the changes are minor and not visible in the mean profile. Turbulent statistics in Figure~\ref{fig:BOMEX_turb_fluxes} are also spatially ($xy$ plane) and temporally (last 3 hrs) averaged and compared to other codes.

\begin{figure}[H]
     \centering
     \includegraphics[width=0.8\textwidth]{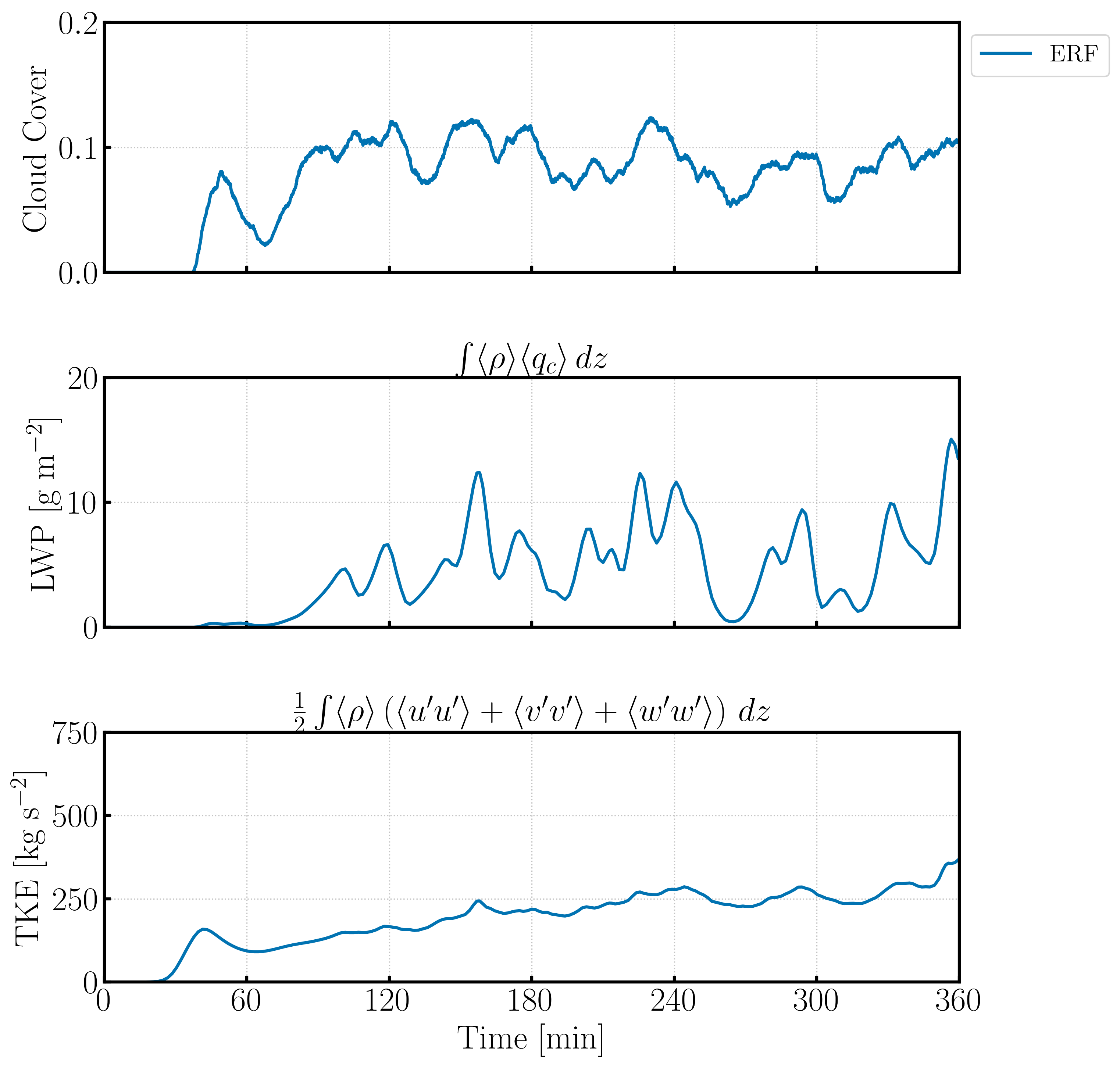}
     \caption{Temporal evolution of the total cloud cover (top), liquid water path (middle), and integrated turbulent kinetic energy (bottom) for the BOMEX case.}
     \label{fig:BOMEX_time_series}
\end{figure}

Temporal evolution of the total cloud cover, domain-integrated liquid water path (LWP), and domain-integrated TKE are also of interest \citep{Siebesma_2003, Sridhar_2022}; see Figure~\ref{fig:BOMEX_time_series}. The total cloud cover plot shows that there is no cloud formation for the first $\sim$ 1/2 hr of spin-up, and after $\sim$ 2 hrs, a statistical steady-state for the the total cloud cover is observed, with a mean of $\approx 0.10$, which is typical for shallow cumulus clouds. The integrated TKE increases steadily with time, and is attributed to mesoscale fluctuations in the horizontal velocities, which increase with time until these fluctuations have the same spatial size as the horizontal domain \citep{Siebesma_2003}.

\subsection{Flow Over Terrain} \label{subsec:terrain}
Until this point, test cases considered herein have employed the standard Cartesian coordinate system without terrain-following or stretched grids. In realistic atmospheric systems, the role of complex terrain cannot be ignored and it is often highly beneficial to use finer resolution near the bottom surface than higher up. To exercise these capabilities, we consider flow over a two-dimensional hill, also known as the \emph{Witch of Agnesi} (WOA) profile. 

\subsubsection{Flow over terrain}

The WOA hill has been widely studied in literature \citep{Z_ngl_2003,Giraldo_2008,Sridhar_2022} and is known to trigger vertically propagating gravity waves. Terrain height is specified as
\begin{equation}
    z(x) = \frac{h_{m} a^2}{\left( x - x_c\right)^2 + a^2},
\end{equation}
where $h_m$ = 1 m and $a = 1000$ m. Freestream conditions are described by a uniform inflow velocity, $U_\infty=10$~m/s and background temperature profile corresponding to a Brunt--V\"ais\"al\"a frequency of $\mathcal{N}=0.01$~1/s. Note that the inverse Froude number, $\frac{\mathcal{N}a}{U_\infty} = 1$ corresponds to the nonhydrostatic regime. The flow is treated as inviscid and dry.

A timestep of 1.0~s is utilized along with a uniform grid resolution of 250~m. The domain is 144~km long and 30~km tall with a 6~km Rayleigh damping layer at the top of the domain. The damping layer depth is approximately equal to one vertical wavelength of the gravity waves, which may be estimated as $\lambda = 2\pi U/\mathcal{N}$ under hydrostatic conditions \citep{Klemp1978_NumericalSimulationHydrostatic}. Analogous damping layers were also included on lateral boundaries with the same damping length and coefficient. The damping coefficient was tuned to $\alpha=0.008$~1/s to minimize wave reflections, falling within the optimal effectiveness range of $2 \le \alpha\lambda/U_\infty \le 5$ \citep{DurranKlemp1983}.

\begin{figure}[htpb!]
\centering
\begin{subfigure}[c]{\textwidth}
        \centering
        \includegraphics[width=0.75\textwidth]{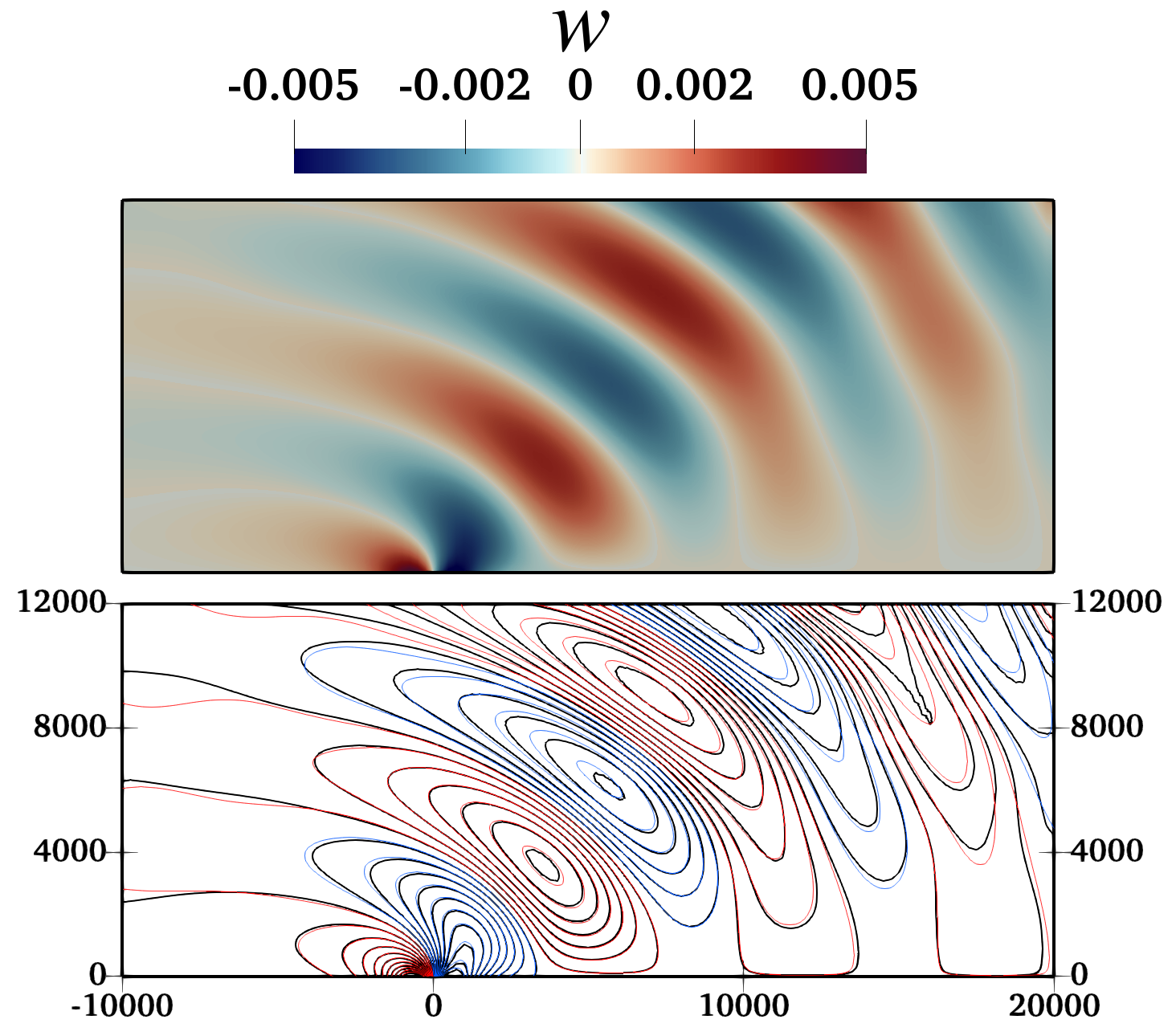}
\end{subfigure}
\caption{Vertical velocity profile (top) and contours (bottom) for the Witch of Agnesi hill at $t$=5 hr; note that only a small subset of the domain is shown in order to focus on the gravity wave propagation. Thick black contours correspond to the ERF simulation while thin red contours correspond to the analytical solution of \citet{smith_linear_1980}. The contours span $\pm 0.005$ m/s with an interval of $0.0005$ m/s.}
\label{fig:woa}
\end{figure}

While the inflow velocity is completely horizontal, the presence of the impenetrable hill forces fluid up and over the terrain, generating a vertical velocity component. Figure~\ref{fig:woa} illustrates the vertical velocity field obtained from ERF as well as the linear analytical solution of \citet{smith_linear_1980}. The presence of gravity waves is readily observed and their vertical structure is consistent with results presented in \citet{Sridhar_2022}; compare Figure~\ref{fig:woa} (top) here to Figure 6b in \citet{Sridhar_2022}. Additionally, strong agreement is observed between ERF and the linear analytical solution, but the differences grow slightly as one moves upward and downstream of the hill.




\section{Software, Parallelism and Performance} \label{sec:software}

\subsection{AMReX}

ERF is built on AMReX \citep{AMReX:IJHPCA, AMReX:IJHPCA2,AMReXweb}, an open-source C++--based software framework that supports the development of structured mesh algorithms for solving systems of partial differential equations, with options for adaptive mesh refinement, on machines from laptops to exascale architectures. AMReX was developed in the U.S. Department of Energy (DOE)’s Exascale Computing Project, receives ongoing support from the DOE OFfice of Science and is now a member project of the High Performance Software Foundation under the umbrella of the Linux Foundation. 

AMReX uses an MPI+X model of hierarchical parallelism where blocks of data are distributed across MPI ranks (typically across multiple nodes).  Fine-grained parallelism at the node level (X) is achieved using OpenMP with tiling for CPU-only machines, or CUDA, HIP or SYCL for NVIDIA, AMD or Intel GPUs, respectively. AMReX enables the use of GPU-aware MPI when available and provides extensive support for kernel launching on GPU accelerators (using ParallelFor looping constructs and C++ lambda functions) and the effective use of various memory types, including managed, device, and pinned.  Common operations, such as parallel communication and reduction operations, as well as interpolation and averaging operators between levels of refinement, are provided by the AMReX framework.  The default load-balancing strategy uses the space-filling curve approach; knapsack and round-robin are also available through AMReX.  Architecture-specific aspects of the software for GPUs are highly localized within the code, and essentially hidden from the application developer or user.  We note that ERF supports both a cmake and a gmake build system.

In addition to portability across architectures, AMReX provides data structures and iterators that define, allocate and efficiently operate on distributed multi-dimensional arrays and Lagrangian particles. Data at each level are defined on disjoint logically rectangular regions of the domain known as patches (or grids or boxes); we recall from Section \ref{subsec:MeshRefinement} that unlike WRF, AMReX (and therefore ERF) does not require one patch per MPI rank, thus allowing much more general domain decomposition. While domain decomposition that assigns a single patch to each MPI rank is often optimal, especially for single-level calculations, there are multilevel cases where we can more efficiently achieve the desired resolution in specific regions by decomposing the region needing refinement into more patches than processors.   The optimal decomposition is problem-specific and ERF provides flexibility in finding the computational ``sweet spot.''

\subsubsection{I/O, Checkpoint/Restart and Visualization}

ERF simulations can be initialized with {\tt metgrid} or {\tt wrfinput} and {\tt wrfbdy} files written in NetCDF format, and users have the option to specify the writing of plotfiles in NetCDF format as well. However, AMReX's native format is more efficient and supported by many third-party visualization tools---e.g., Paraview, VisIt, yt and other python tools.
For efficiency all checkpoint/restart files are written in AMReX native format.  This format is a combination of text header files holding the metadata and binary files containing the data itself, stored in a directory structure with data separated by refinement level. The reading and writing of these files can be done asynchronously so that the computation can proceed while the data is being written out.   We also note that posptrocessing tools to convert AMReX native plotfiles to NetCDF files are available.

AMReX has multiple output methodologies to provide efficient I/O across a variety of applications and simulations. First, a static output pattern prints in a pre-determined pattern that eliminates unnecessary overhead and is useful for well-balanced or small simulations. Second, a dynamic output pattern improves write efficiency for complex cases by assigning ranks to coordinate the I/O in a task-like fashion. Finally, asynchronous output utilizes a background thread and data copy for writing, thereby allowing computation to continue uninterrupted. The performance of the different I/O options is discussed in the next section.


\subsection{Performance} \label{sec:perf}
To assess the performance of the ERF code on CPUs and GPUs, strong and weak scaling studies were completed. A canonical ABL LES (see Section 4.2.1) was chosen for the scaling studies, with third-order advection and Smagorinsky turbulence closure. The scaling was performed on the CPU and GPU nodes on Perlmutter, a supercomputer hosted at the National Energy Research Scientific Computing Center (NERSC), which is located at Lawrence Berkeley National Laboratory (Berkeley Lab). A single CPU node consists of 128 AMD EPYC 7763 (Milan) cores while a GPU node has 64 AMD EPYC 7763 (Milan) cores and 4 NVIDIA A100 (Ampere) GPUs. In all simulations the domain has lengths $(L_x, \, L_y, \, L_z)$ = (2048, 2048, 1024) m.

\begin{figure}
    \centering
    \includegraphics[width=\textwidth]{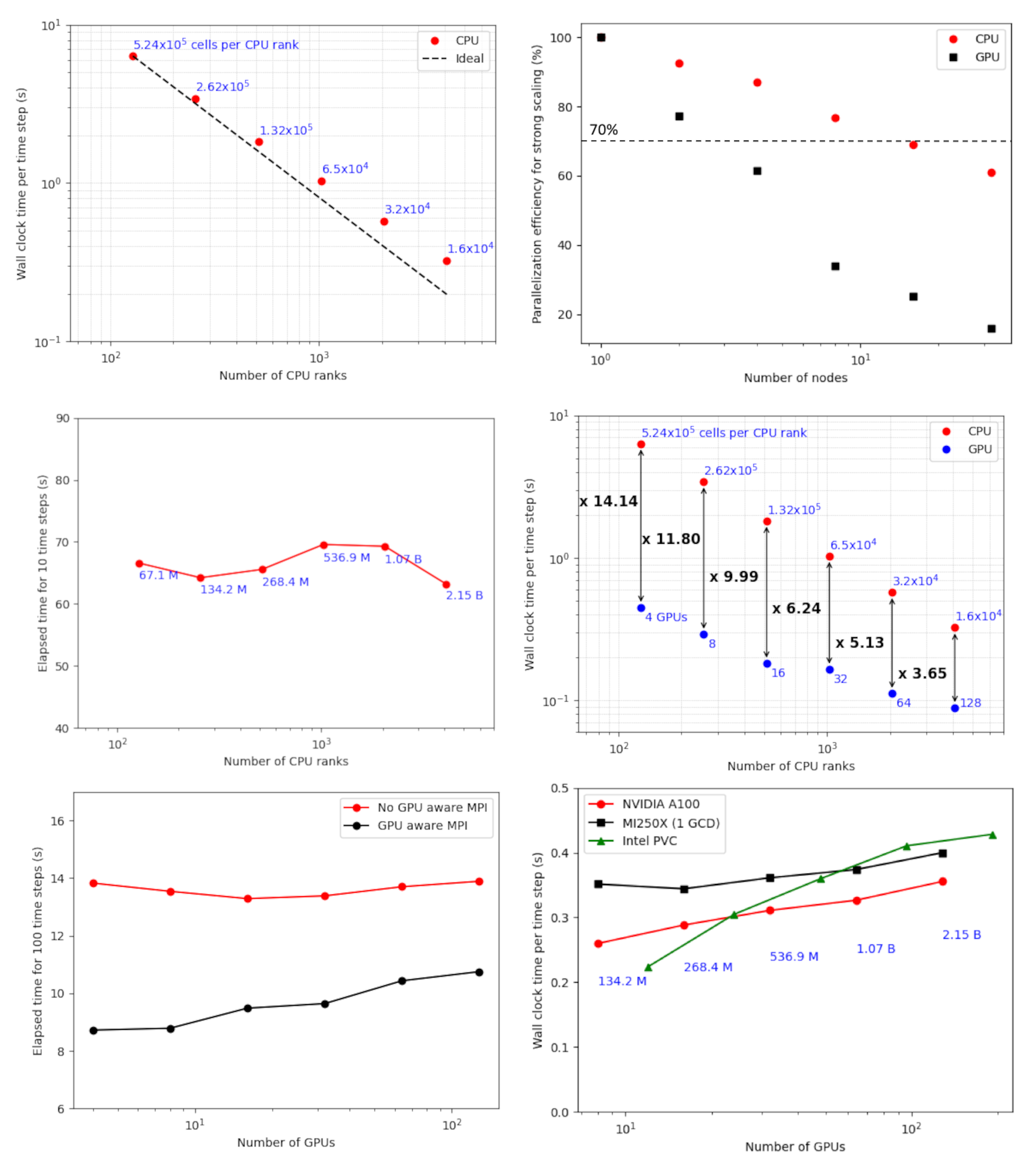}
    \caption{(top left) Strong scaling on CPUs. The number of mesh cells per rank is shown in blue. (top right) Parallelization efficiency for strong scaling. (middle left) Weak scaling on CPUs. The total number of mesh cells is shown in blue. (middle right) Comparison of timings on CPU and GPU showing the speed-up factor. We compare a CPU node with 128 ranks to a GPU node with 4 ranks, so there are 32x more points per GPU than per CPU core.  Points on the same vertical line represent the same number of nodes. (bottom left) Weak scaling on GPUs with and without GPU-aware MPI. (bottom right) Weak scaling on GPUs: Perlmutter (NVIDIA A100), Frontier (AMD MI250X (1 GCD)) and Aurora (Intel Data Center GPU Max Series (PVC)). \label{fig:performance}}
\end{figure}

For the strong scaling study, a fixed mesh size of $(N_x, \, N_y, \, N_z)$ = (512, 512, 256) is employed and a 128 core simulation is taken as the reference. Subsequent simulations were completed by doubling the number of cores until 4096. For CPU-only simulations, the observed strong scaling timings are illustrated in Figure~\ref{fig:performance} (top left) while the parallelization efficiency, defined as
\begin{eqnarray*}
E = \Bigg(\frac{T}{T_\text{128}}\Bigg)\Bigg(\frac{\text{128}}{N}\Bigg) \times 100 \%,
\end{eqnarray*}
is shown in Figure~\ref{fig:performance} (top right). In the above, $N$ is the number of cores and $T$ is the time taken per time step. With CPUs, at 2048 cores, and $\sim 32^3$ cells per rank, the parallel efficiency is 69\%, and for GPUs (NVIDIA A100), the parallelization efficiency was 70\% when using 3 nodes (12 GPUs) (6 million cells per GPU).

On GPU nodes, the number of CPU ranks and GPUs are the same for each run (ie. when running with 8 GPU nodes ie., 32 GPUs, the number of CPU ranks is 32 as well---i.e., each rank offloads its work to a single GPU. This GPU run with 8 GPU nodes ie., 32 GPUs, is compared to a CPU run with 8 CPU nodes ie. 1024 CPU ranks). Therefore, the speed-up presented here between GPU and CPU is per node. Speed-ups of 5--15$\times$ are achieved up to 16 GPU nodes; see Figure~\ref{fig:performance} (middle right).

A weak scaling test was performed on CPUs with a mesh size of $512\times512\times256 = 67.1$ million, on 1 node (128 MPI ranks), and the number of cells was progressively scaled to $4096\times2048\times256 = 2.1$ billion, on 32 nodes (4096 MPI ranks). Excellent weak scaling is shown by the nearly constant timings, for 10 timesteps, in Figure~\ref{fig:performance} (middle left).

A weak scaling test was also performed on GPUs with a mesh size of $256\times256\times512$ on 1 GPU node, and the number of cells was progressively scaled to $2048\times1024\times512$ on 32 nodes. The total elapsed time for 100 iterations with and without GPU-aware-MPI is shown in Figure~\ref{fig:performance} (bottom). The nearly constant timings show excellent weak scaling and the benefits of GPU-aware-MPI are clearly observed with the 25-35\% speed-up.

I/O timings are dependent on many factors including the specifics of the architecture and file system as well as the size and load balance of the problem.  We present two sets of timings here to give a ballpark idea of relative costs of the I/O options. 
For a simple CPU-only calculation on 4 MPI ranks with 512x128x1 mesh points, writing plotfiles with 10 variables every 10 steps increased the runtime by less than 1\% using the AMReX format and by over 7\% using the NetCDF format.  In a CPU-only comparison with WRF for a domain with 12.9 million grid cells, we found that the walltime for WRF to write a NetCDF file was roughly 9 seconds while the wallclock time for ERF to write a plotfile in native AMReX format was roughly 0.6 seconds. When running with GPUs, using asynchronous I/O with the native AMReX format can further reduce the write time.

\begin{figure}
    \centering
    \includegraphics[width=0.5\textwidth]{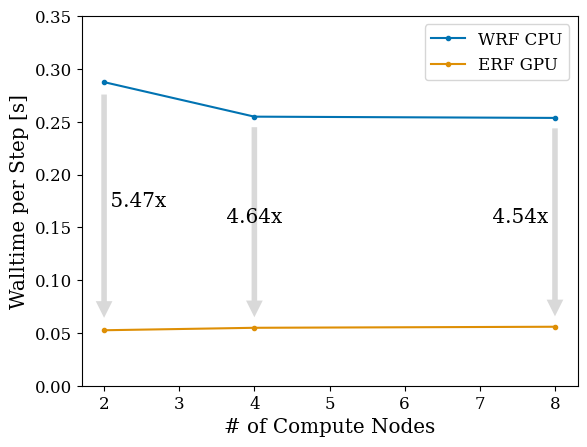}
\caption{Timing comparison between ERF (using GPUs) and WRF (CPU only) on the Perlmutter supercomputer for a 3D squall line simulation with Kessler microphysics.}
\label{fig:erfvswrf}
\end{figure}

Finally, to evaluate the runtime of ERF relative to WRF for a problem with simple cloud microphysics, we performed
a set of 3-D squall line simulations (a larger-scale variant of the problem presented in Section~\ref{subsub:squall}). 
ERF and WRF simulations were run on 2 Perlmutter nodes for a problem size of $401 \times 401 \times 80 = 12.86$ million cells, on 4 nodes with twice as many cells, and on 8 nodes with four times the problem size. In this study, ERF ran on the Perlmutter GPU nodes (using all 4 GPUs per node) and WRF ran on the Perlmutter CPU nodes (using all 64 MPI ranks per node).
Wallclock time per computational time step for ERF and WRF is illustrated in Figure~\ref{fig:erfvswrf}, demonstrating a roughly 5x speedup through leveraging GPU acceleration.

\section{Conclusions and Future Work} \label{sec:conclusion}

Since the introduction of ERF \citep{ERF_joss}, significant development has occurred that includes core modeling capabilities, Lagrangian particles, and adaptive mesh refinement. These features are, by design, highly modular and allow users to dynamically select between compressible and anelastic formulations and to choose the relevant physics models (e.g., LES vs. PBL, or dry vs. moist) for a given problem of interest. Thus ERF is well-suited for both exploratory studies and production runs of regional atmospheric flows. ERF's use of the state-of-the-art and well-supported AMReX framework ensures it will continue to run efficiently as architectures and operating systems evolve; AMReX shields ERF developers and users from many of the code changes necessary to operate on new architectures.

This paper provides a detailed overview of the ERF software and its current capabilities. A variety of test cases 
were presented to compare ERF with benchmark data. For each problem considered here, ERF demonstrated the ability to replicate established results. These findings represent the first steps towards establishing ERF as a modern regional weather modeling tool and an attractive multi-GPU-capable replacement for WRF (or similar atmospheric simulation codes with limited GPU applicability).

Future work includes continued development and validation of ERF. 
Over the next year, validation studies will focus on real-world cases with available observational data, including but not limited to simulation of flow over steep complex terrain as well as simulation of extreme events such as recent hurricanes.  New capability development will include the generalization of meshing strategies
to allow an embedded boundary (``cut cell'') representation of terrain and/or urban structures as well as use of the particle data structures to develop and test Lagrangian microphysics models.  

Looking forward, the plan for ERF is to leverage existing modules built on years of experience by practitioners who are experts in their domains. Rather than re-implementing existing column physics modules such as SHOC, P3, and RRTMGP, which exist in the E3SM GitHub repository~\citep{E3SM}, we plan to provide interfaces in ERF to call the versions of these routines that have already been ported to C++ and can run on GPUs through the use of Kokkos~\citep{kokkos}.  Similarly, we will also leverage the existing Noah-MP \citep{he_community_2023} and SLM  \citep{lee_simplified_2015} land models. To this end, a rewrite of SLM has been incorporated into ERF along with interfaces to RRTMGP and Noah-MP. However, the aforementioned physics are currently undergoing testing and verification.

Creating interfaces to existing physics models serves two purposes. First, it clearly increases the fidelity of ERF to model atmospheric phenomena.
Second, it enables ERF to serve as a testbed to facilitate the rapid development and testing of new features in the column physics modules themselves.  Currently, for development purposes, these models are often used either in single-column mode or deployed in E3SM itself, with no intermediate options.   ERF can provide an easy-to-use intermediate testing framework that will allow fast development and testing of column physics models in a limited-area testbed that fills the gap between single-column and global.

In addition to calling additional physics models as described above, we are establishing strategies for run-time coupling with WW3~\citep{WW3}, a wave model framework that solves the random phase spectral action density balance equation for wavenumber-direction spectra, and with REMORA~\citep{REMORA}, a new AMReX-based ocean model built to reproduce the ROMS~\citep{ROMS} regional ocean modeling capability on modern architectures.  This three-way coupling will be analogous to COAWST~\citep{COAWST} in which WRF, ROMS and SWAN or WW3 are three-way coupled.  There is also ongoing work to couple ERF to AMR-Wind.

Finally, efficiently exploiting modern and future architectures and using performant implementations of high-fidelity models is not in itself sufficient to address the large problems spanning multiple length and time scales that we target. Development of surrogate models for various components of the overall model require extensive training data and ways to integrate the simulation code, written in C++, with AI training approaches which are typically written in Python. The Python language bindings of pyAMReX~\citep{AMReX:IJHPCA2} could enable ERF to fit easily into AI/ML workflows.


\section{Data Availability Statement}

The ERF model is completely open-source and is released under a modified BSD license.  It is available at {\url{https://github.com/erf-model/ERF}}.  Documentation of the ERF code can be found at {\url{https://erf.readthedocs.io}}. The github hash and inputs files for all of the results presented here are available at
{\url{https://github.com/erf-model/validation/tree/main/JAMES\_Paper}}. The GitHub code will be linked to Zenodo and provided with the DOI citation in the references list and in-text citation upon acceptance.

\acknowledgments
Funding for this work was provided by the U.S. Department of Energy
Office of Energy Efficiency and Renewable Energy Wind Energy Technologies Office.
We acknowledge as well as the help of the AMReX team
in developing and supporting new AMReX features needed by ERF.
AMReX is supported in part by the U.S. Department of Energy (DOE) Office of Advanced Scientific Computing Research (ASCR) via the Scientific Discovery through Advanced Computing (SciDAC) program FASTMath Institute.
The work at LBNL was supported by the U.S. Department of Energy
under contract No. DE-AC02-05CH11231.
The work at LLNL was supported by the U.S. Department of Energy
under contract No. DE-AC52-07NA27344.
The contribution of Weiqun Zhang was supported in part by the U.S. Department of Energy, Office of Science, Office of Advanced Scientific Computing Research, Scientific Discovery through Advanced Computing (SciDAC) Program through the FASTMath Institute under Contract No. DE-AC02-05CH11231 at Lawrence Berkeley National Laboratory.  
Funding for the implementation of actuator disk models came from the FLOWMAS project funded by the U.S. Department of Energy, Office of Science Energy Earthshot Initiative.
The contribution of Branko Kosovic was supported by the National Center for Atmospheric Research, which is a major facility sponsored by the National Science Foundation under Cooperative Agreement No. 1852977.
This work was authored in part by the National Renewable Energy Laboratory for the U.S. Department of Energy (DOE) under Contract No. DE-AC36-08GO28308. The views expressed in the article do not necessarily represent the views of the DOE or the U.S. Government. The U.S. Government retains and the publisher, by accepting the article for publication, acknowledges that the U.S. Government retains a nonexclusive, paid-up, irrevocable, worldwide license to publish or reproduce the published form of this work, or allow others to do so, for U.S. Government purposes.

In addition, our thanks to  Emily Bercos-Hickey, Matthew Churchfield, Harish Gopalan, Soonpil Kang and Kyle Pressel for their constructive feedback on earlier drafts of this manuscript. Furthermore, we would like to thank professor Frank Giraldo for his constructive feedback on the linear solutions for flow past terrain.

A portion of this code development was performed using computational resources sponsored by the U.S. Department of Energy's Office of Energy Efficiency and Renewable Energy and located
at the National Renewable Energy Laboratory. This development also used resources of the
National Energy Research Scientific Computing Center (NERSC),
a U.S. Department of Energy Office of Science User Facility located at
Lawrence Berkeley National Laboratory.

\appendix

\section{Terrain Coordinates and Map Scale Factors} \label{app:A}

Consider two coordinate systems that correspond to a terrain-following grid, $\bm{X}$, and a flat cartesian grid, $\bm{Z}$, with axes given by 
\begin{align}
 \bm{X} &= \left[ x \; y \; z \right]^{\intercal}, \quad \quad   
 \bm{\Xi} = \left[ \xi \; \eta \; \zeta \right]^{\intercal},
\end{align}
and
\begin{equation}
x = \xi, \quad \quad y = \eta, \quad \quad z =  h \left(\xi, \, \eta, \, \zeta \right).
\end{equation}
Only the vertical coordinate in the physical domain is deformed by the terrain-fitting.
To account for isotropic lateral grid stretching as represented by ``map factors" $m_x = m_y = m$ as in WRF, we augment the coordinate transform above with stretching in the lateral directions only.
These combined transformations yield the following Jacobian, $\bar{\bm{J}}$, and inverse Jacobian, $\bar{\bm{T}}$, matrices
\begin{equation}
    \bar{\bm{J}}  = \begin{bmatrix}
    \frac{1}{m} & 0 & 0 \\
    0 & \frac{1}{m} & 0\\
   h_{\xi} &  h_{\eta} & h_{\zeta} \\
    \end{bmatrix}, \quad \quad 
     \bar{\bm{T}} =  \bm{J}^{-1} =  \frac{m^2}{h_{\zeta}} \begin{bmatrix}
    \frac{h_{\zeta}}{m} & 0 & 0 \\
    0 & \frac{h_{\zeta}}{m} & 0\\
   -\frac{h_{\xi}}{m} &  -\frac{h_{\eta}}{m} & \frac{1}{m^2} \\
  \end{bmatrix} 
  = 
   \begin{bmatrix}
    m & 0 & 0 \\
    0 & m & 0\\
   -\frac{h_{\xi}}{h_\zeta}m &  -\frac{h_{\eta}}{h_\zeta}m & \frac{1}{h_\zeta} \\
    \end{bmatrix}. \label{eq:jacobian}
\end{equation}
In the above, $J = \left| \bar{\bm{J}} \right |=  h_{\zeta} / m^2$ is the Jacobian determinant. To explicitly close the governing equations in terrain-following coordinates (Eqs.~\ref{eq:con}--\ref{eq:precip}), we provide relations for the gradient of a scalar ($f$) and divergence of a vector ($\bm{F}$):
\begin{align}
    \nabla_{\bm{X}} f &= \bar{\bm{T}}^{\intercal} \nabla_{\bm{Z}} f, \label{eq:terraingrad}\\
    \nabla_{\bm{X}} \cdot \left( \bm{F} \right) &= \frac{1}{J} \nabla_{\bm{Z}} \cdot \left( J  \bar{\bm{T}} \bm{F}\right). \label{eq:terraindiv}
\end{align}
Vector rotation of the fluid velocity from Eq.~\ref{eq:terraindiv} yields $J  \bar{\bm{T}} \bm{u} = \left[h_{\zeta}u/m, \, h_{\zeta}v/m, \, \omega/m^2  \right]^{\intercal}$, where $\omega = w -h_{\xi} u m - h_{\eta} v m$ is the vertical velocity that is normal to the top/bottom faces of the grid cells.

\section{Terrain Metrics} \label{app:B}

In practice, we define the mapping $z =  h (\xi, \, \eta, \, \zeta)$ at nodes (cell corners), and we define the two-dimensional arrays of map factors at cell centers and faces.
The metric terms $h_\xi, h_\eta$ and $h_\zeta$ are defined as gradients formed from the nodal values of $h$. Here we write the x-momentum equation explicitly:
\begin{align}
\frac{\partial \rho_d u }{\partial t} &= - \frac{m^2}{h_{\zeta}} \left(
\frac{\partial (\rho_d u u h_{\zeta} m^{-1})}{\partial \xi} +
\frac{\partial (\rho_d v u h_{\zeta} m^{-1})}{\partial \eta} \right)
- m^2\frac{\partial \left(\rho_d \omega u m^{-2} \right)}{\partial \zeta} - m\left(p_\xi - p_\zeta \frac{h_\zeta}{h_\xi} \right). \label{eq:contex}
\end{align}

As discussed by \citet{Klemp:2003}, the $h_\xi$ and $h_\eta$ metrics should be computed in a manner that satisfies a discrete cancellation property. More specifically, the discretizations employed for $h_{\xi}; \, h_{\eta}$ should be consistent with those employed for the pressure gradient and advection operator. We note that in ERF a staggered $2^{\rm nd}$ order discretization is applied to the pressure gradient while the advection operator defaults to $3^{\rm rd}$ order upwind, which is a centered $4^{\rm th}$ scheme with an upwind factor. Therefore, the metric discretizations should be consistent with the aforementioned discretizations.

For the pressure gradient $p_x = p_\xi - p_\zeta \left(h_\zeta / h_\xi \right)$ at the $(i+\half,j,k)$ face, we write 
\begin{align}
 (p_x)_{i+\frac{1}{2},j,k} &= \frac{1}{\Delta \xi} (p_{i+1,j,k} - p_{i,j,k}) - \frac{1}{2} (h_\xi)_{i+\frac{1}{2},j,k} \left( \left(\frac{p_\zeta}{h_\zeta}\right)_{i+1,j,k} + 
                 \left(\frac{p_\zeta}{h_\zeta}\right)_{i,j,k} \right),                
\end{align}
where the vertical pressure gradient is computed as 
\begin{align}
\left(\frac{p_\zeta}{h_\zeta}\right)_{i,j,k} &= \frac{p_{i,j,k+1} - p_{i,j,k-1}}{z_{i,j,k+1} - z_{i,j,k-1}},
\end{align}
with $h_\xi$ defined at $(i+\half,j,k)$ face by 
\begin{align}
    \left(h_\xi\right)_{i+\frac{1}{2},j,k} & = \frac{z_{i+1,j,k} - z_{i,j,k}}{\Delta \xi}
\end{align}
where $z_{i,j,k}$ is obtained from averaging the height at the 8 surrounding nodes.

Similarly, $\omega = w - h_\xi um - h_\eta vm$ is defined at the $(i,j,k+\half)$ face by 
\begin{align} \omega_{i,j,k+\frac{1}{2}} &=  w_{i,j,k+\frac{1}{2}} \\
&- \frac{1}{4} \left[ (h_\xi)_{i,j,k+\frac{1}{2}} \left( 
(um)_{i+\frac{1}{2},j,k} + (um)_{i-\frac{1}{2},j,k} + 
 (um)_{i+\frac{1}{2},j,k+1} + (um)_{i-\frac{1}{2},j,k+1} \right) \right. \nonumber\\
&\left. \quad \;\;+  (h_\eta)_{i,j,k+\frac{1}{2}} \left( (vm)_{i,j+\frac{1}{2},k} + (vm)_{i,j-\frac{1}{2},k} + (vm)_{i,j+\frac{1}{2},k+1} + (vm)_{i,j-\frac{1}{2},k+1} \right) \right], \nonumber 
\end{align}
where $h_\xi$ and  $h_\eta$ are defined at the $(i,j,k+\half)$ face by
\begin{align}
    (h_\xi)_{i,j,k+\frac{1}{2}} = \frac{z_{i-2,j,k+\frac{1}{2}} - z_{i+2,j,k+\frac{1}{2}} + 8 \left( z_{i+1,j,k+\frac{1}{2}} - z_{i-1,j,k+\frac{1}{2}}\right)}{12 \Delta \xi},
\end{align}
and $z_{i,j,k+\frac{1}{2}}$ is obtained from averaging the 4 nodes surrounding the $(i,j,k+\half)$ face.

\begin{figure}[htpb!]
\centering
\begin{subfigure}[c]{\textwidth}
        \centering
        \includegraphics[width=0.75\textwidth]{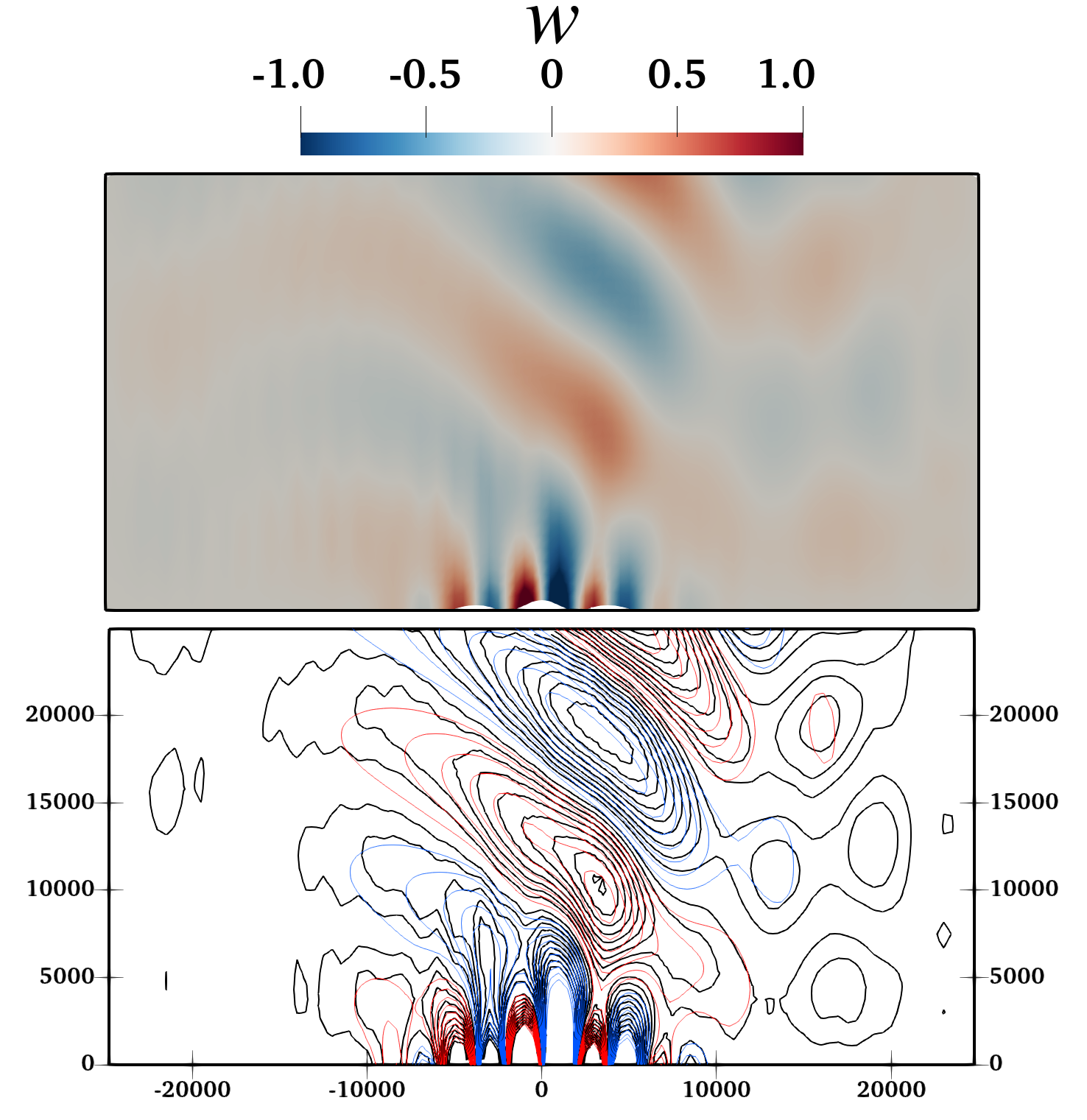}
\end{subfigure}
\caption{Vertical velocity profile (top) and contours (bottom) for the Sch\"{a}r mountain at $t$=5 hr. Thick black contours correspond to the ERF simulation while thin red/blue contours correspond to the analytical solution of \citet{Klemp_2015}. The contours span $\pm 0.5$ m/s with an interval of $0.05$ m/s.  }
\label{fig:schar}
\end{figure}

To test the consistency of the metric terms provided above, we consider the mountain case of \citet{Schar_2002}. The conditions employed for the Sch\"{a}r mountain follow those described in \citet{Klemp:2003}. We note that the centered $4^{\rm th}$ order advection scheme was employed for consistency with $\omega$ and a time step of $\Delta t = 1$ s was utilized. The vertical velocity profile and comparison with the linear analytical solution of \citet{Klemp_2015} are illustrated in Figure~\ref{fig:schar}. in Figure~\ref{fig:schar} (bottom), as one moves up from the mountain, a single closed set of positive contour lines are observed, which is a key feature of the discrete cancellation property; see Figure 1a vs 1c of \cite{Klemp:2003}.

\section{Acoustic Substepping} \label{app:C}

In the acoustic substepping,
we first define $(\rho_d \mathbf{u})^t, (\rho_d \theta_{d})^t,$ $\rho_{d}^t$, $q_v^t$ and $q_c^t$ as the solution at the most recent RK stage; these are held fixed during the acoustic substepping.  We then 
define solution variables $(\Up, \Vp, \Wp) = (\rho_d \mathbf{u}) - (\rho_d \mathbf{u})^t,$ $\Tp = (\rho_d \theta_{d}) - (\rho_d \theta_{d})^t,$ and $\Rp = \rho_{d} - \rho_{d}^t$ and the change over an acoustic substep $\Delta U^{\prime \prime} = \Uppt - \Upp$. 

In each substep, to evolve the solution by $\delta \tau$, where $\delta \tau$ is the substepping time step, we solve
\begin{eqnarray*}
\Delta U^{\prime \prime} &=&  \delta \tau \left(
              -\frac{\gamma R_d \pi^t m}{1 + q_v^t + q_c^t} \left( \frac{\partial \Theta^{\prime \prime, \tau}}{\partial \xi} - \frac{h_\xi}{h_\zeta} \frac{\partial \Theta^{\prime \prime, \tau}}{\partial \zeta}\right) + R^t_U \right) \\ \\
  \Delta V^{\prime \prime} &=&  \delta \tau \left(
              -\frac{\gamma R_d \pi^t m}{1 + q_v^t + q_c^t} \left( \frac{\partial \Theta^{\prime \prime, \tau}}{\partial \eta} - \frac{h_\eta}{h_\zeta} \frac{\partial \Theta^{\prime \prime, \tau}}{\partial \zeta}\right) + R^t_V \right) \\ \\
\Delta W^{\prime \prime} &=&  \delta \tau \left(
            -\frac{\gamma R_d \pi^t}{1 + q_v^t + q_c^t}  \frac{\partial (\beta_1 \Tpp + \beta_2 \Tppt ) }{h_\zeta \partial \zeta} \right. \\
            && \left. + g \overline{\rho} \frac{\rho_d}{\rho_m} \frac{R_d}{c_v} \frac{\pi^t}{\overline{\pi}}
             \frac{ (\beta_1 \Tpp  + \beta_2 \Tppt )}{\Theta^t}
            - g (\beta_1 \Rpp + \beta_2 \Rppt ) + R^t_W \right) \\ \\ 
\Delta \Theta^{\prime \prime} &=&   \delta \tau \left(
          -\frac{m^2}{h_{\zeta}} \frac{\partial \left(h_\zeta \Uppt \theta_{d}^t m^{-1}\right)}{\partial \xi}
          -\frac{m^2}{h_{\zeta}} \frac{\partial \left(h_\zeta \Vppt\theta_{d}^t m^{-1}\right)}{\partial \eta} \right. \\
          && \left . -\frac{1}{h_{\zeta}} \frac{\partial \left(\left( \beta_1 \Omega^{\prime \prime, \tau} + 
                             \beta_2 \Omega^{\prime \prime, \tau + \delta \tau} \right) \theta_{d}^t \right)}{\partial \zeta} +  R^t_{\Theta} \right) \\ \\
\Delta \rho^{\prime \prime} &=&   \delta \tau \left(
          - \frac{m^2}{h_{\zeta}} \frac{\partial \left(h_\zeta \Uppt m^{-1} \right)}{\partial \xi}
          - \frac{m^2}{h_{\zeta}} \frac{\partial \left(h_\zeta \Vppt m^{-1} \right)}{\partial \eta} \right. \\
          && \left. - \frac{1}{h_{\zeta}} \frac{\partial \left(\beta_1 \Omega^{\prime \prime, \tau} +
                            \beta_2 \Omega^{\prime \prime, \tau + \delta \tau} \right)}{\partial \zeta} +  R^t_{\rho} \right)
\end{eqnarray*}
where $g$ is the positive value of the gravitational acceleration, $\pi$ is the Exner function, and $\Omega = W -h_\xi U m - h_\eta V m$ is the contravariant vertical momenta.  $R_U, R_V, R_W, R_\Theta$ and $R_\rho$ hold the right-hand-sides of Equations~\ref{eq:con}--\ref{eq:theta}.

%
Following Equation (17) in \citet{klemp_conservative_2007}, $\beta_1 = (1 - \beta_s)/2$, $\beta_2 = (1 + \beta_s) / 2.$  The overbar notation, again following the notation in  \citet{klemp_conservative_2007}, can be expressed for any variable $\phi$ as 
$\overline{\phi}^\tau = \beta_2 \phi^{\tau + \Delta \tau} + \beta_1 \phi^{\tau}$. Thus, the $\beta_s$ controls the degree of implicitness; as in \citet{klemp_conservative_2007} we set $\beta_s = 0.1$.

Once the acoustic substepping is completed for a particular RK stage, the perturbational solution is added to the solution at the most recent RK stage. 

\section{Simulation Conditions} \label{app:D}

Table~\ref{tab:BigObnoxiousThing} summarizes the simulation configurations for the cases presented in Section~\ref{sec:vandv}. Furthermore, for the sake of readability, we reiterate the table description below.

The domain extents describe the height of the domain in the 1-D single-column model (SCM) case; the length and height of the domain for the 2-D cases; and the streamwise, lateral, and vertical extents of the domain for the 3-D cases. Parenthetical quantities are derived from other inputs and have been included for completeness. Initial fields are either uniform or vary with height based on an input sounding. A perturbation field is superimposed on the initial fields to model the specific mesoscale problem of interest or to help initiate turbulence in the microscale. For the canonical, non-neutral ABL simulations, MOST specifies the surface kinematic heat flux for the BOMEX case and the surface temperature for the GABLS cases. Both Smagorinsky and Deardorff 1.5 order TKE turbulence closures have been utilized; with Deardorff, the model coefficients follow \citep{Moeng1984}. 
Several cases used Rayleigh damping to attenuate waves in the free atmosphere.
The damping strength, an inverse timescale, was ramped using a sine-squared function from 0 at the bottom of the damping layer to the specified value at the top.

\begin{sidewaystable}
\centering
\renewcommand{\arraystretch}{1.3} 
\resizebox{\textwidth}{!}{%
\Huge 
\begin{tabular}{|L{2.75in}|C{2in}|C{2in}|C{3in}|C{1in}|C{2in}|C{3.375in}|c|C{4in}|c|C{2.5in}|C{1.75in}|C{2.25in}|C{2.625in}|C{2.75in}|}
\hline
 & \multicolumn{3}{|c|}{\textbf{Domain}} &  & \multicolumn{2}{|c|}{\textbf{Initial Conditions}} & \multicolumn{3}{|c|}{\textbf{Boundary Conditions}} &  &  &  &  &  \\
\cline{2-4} \cline{6-7} \cline{8-10} 
\textbf{Case} & \textbf{Domain Extent(s)} [km] & \textbf{Number of Grid Cells} & \textbf{Grid Spacing} [m] & \textbf{Time Step} [s] & \textbf{Type} & \textbf{Perturbations} & \textbf{Lateral} & \textbf{Bottom} & \textbf{Top} & \textbf{Turbulence Closure} & \textbf{Microphysics} & \textbf{Geostrophic Wind} [m/s] & \textbf{Coriolis Frequency} [s$^{-1}$], \textbf{Latitude} [$^\circ$] & \textbf{Rayleigh Damping Layer} Depth, Strength: Applied Field(s) \\
\hline
Density Current & 25.6, 6.4 & 256, 64 & (100, 100) & 1.0 & Uniform & Cold bubble & Symmetry, Outflow & Slip Wall & Slip Wall & — & — & — & — & — \\ \hline
Bubble Rise & 20., 10. & 200, 100 & (100., 100.) & 0.5 & Uniform & Warm bubble & Slip Wall & Slip Wall & Slip Wall & — & None / Kessler (no precip.) & — & — & — \\
\hline
Squall Line & 150., 24. & 1500, 240 & (100., 100.) & 0.25 & Uniform & Warm bubble, wind shear & Open & Slip Wall & Outflow & Smagorinsky ($C_s=0.25$) & Kessler & — & — & — \\
\hline
Supercell & 150., 100., 24. & 600, 400, 96 & (250., 250., 250.) & 0.5 & Uniform & Warm bubble, wind shear & Periodic, Open & Slip Wall & Outflow & Smagorinsky ($C_s=0.25$) & Kessler & — & — & — \\
\hline
Neutral ABL LES, grid aspect ratio=4 & 2.4, 2.4, 2.4 & 160, 160, 210 & (15., 15., 5.) & 0.2 & Sounding & Divergence-free $\Delta u, \Delta v = \pm0.65$ m~s$^{-1}$ & Periodic & MOST: $z_0 = 0.05$ m & Slip Wall & Deardorff & — & [6.5, 0.] & (0.803$\times10^{-4}$), 33.5$^\circ$ & 400~m, 0.003~s$^{-1}$: $u$, $v$, $w$, $\theta$ \\
\hline
Neutral ABL LES, isotropic grid & 2.4, 2.4, 2.4 & 320, 320, 140 & (7.5, 7.5, 7.5) & 0.1 & Sounding & Divergence-free $\Delta u, \Delta v = \pm0.65$ m/s & Periodic & MOST: $z_0 = 0.05$ m & Slip Wall & Deardorff & — & [6.5, 0.] & (0.803$\times10^{-4}$), 33.5$^\circ$ & 400~m, 0.003~s$^{-1}$: $u$, $v$, $w$, $\theta$ \\
\hline
GABLS SCM & 0.4 & 64 & (6.25) & 1.0 & Sounding & — & Periodic & MOST: $z_0=0.1$ m, 0.25 K~hr$^{-1}$ cooling & Slip Wall & MYNN Level 2.5 & — & [8., 0.] & $1.39\times10^{-4}$, $73.^\circ$ & — \\
\hline
GABLS LES & 0.4, 0.4, 0.4 & 128, 128, 128 & (3.125, 3.125, 3.125) & 0.05 & Sounding & Random $\Delta\rho\theta = \pm0.1$~K & Periodic & MOST: $z_0=0.1$ m, 0.25 K~hr$^{-1}$ cooling & Slip Wall & Smagorinsky ($C_s=0.17$) / Deardorff & — & [8., 0.] & $1.39\times10^{-4}$, $73.^\circ$ & 100~m, 0.2~s$^{-1}$: $w$ \\
\hline
BOMEX & 6.4, 6.4, 4.0 & 64, 64, 100 & (100., 100., 40.) & 0.3 & Sounding & Random $\Delta\rho\theta=\pm0.1$~K, $\delta q_v = \pm0.025$~g~kg$^{-1}$ & Periodic & MOST: $z_0=0.1$ m, $u_*=0.28$ m~s$^{-1}$, $\langle w'\theta' \rangle = 8\times10^{-3}$~K~m~s$^{-1}$, $\langle w'q_t' \rangle = 5.2\times10^{-5}$~m~s$^{-1}$ & Slip Wall & Smagorinsky ($C_s=0.17$) & Kessler (no precip.) & Time-varying profile & $0.376\times10^{-4}$, ($15.^\circ$) & — \\
\hline
Flow Over Terrain & 144., 30. & 288, 600 & 500., initial $\Delta z=0.1$ w/ stretching ratio=1.05 & 0.06 & Uniform & — & Inflow, Outflow & Slip Wall & Slip Wall & — & — & — & — & 5~km, 0.2~s$^{-1}$: $w$ \\
\hline
\end{tabular}%
}
\caption{Simulation configurations for the verification and validation cases presented in Section~\ref{sec:vandv}.
\label{tab:BigObnoxiousThing}}
\end{sidewaystable}

\clearpage

\bibliography{ERF.bib}

\end{document}